\documentclass[10pt]{article}

\usepackage{amsfonts,amssymb,amsmath,mathrsfs,latexsym,bbm,dsfont,amsthm}
\usepackage{color}
\usepackage{graphicx}
\usepackage{pstricks,pstricks-add}
\usepackage{pst-plot}
\usepackage[scriptsize,nooneline,hang]{caption}
\usepackage[hang,nooneline,scriptsize]{subfigure}
\usepackage{rotating}
\usepackage[all]{xy}

\usepackage[abs]{overpic}

\definecolor{lightgrey}{rgb}{0.95,0.95,0.95}\definecolor{grey}{gray}{0.7}

\newcommand{\tr}{\tilde{r}}

\newcommand{\Dr}{\Delta_r}
\newcommand{\tDr}{\tilde{\Delta}_r}
\newcommand{\half}{\frac{1}{2}}
\newcommand{\tL}{\tilde{L}}

\newcommand{\De}{\Delta_{\rm q}}\newcommand{\Des}{\Delta^2_{\rm q}}
\newcommand{\Dg}{\Delta_{\rm g}}\newcommand{\Dgs}{\Delta^2_{\rm g}}

\newcommand{\tq}{\tilde{Q}}
\newcommand{\tg}{\tilde{G}}

\begin{document}

\title{Geodesics of electrically and magnetically charged test particles 
in the Reissner-Nordstr\"om space-time: analytical solutions}

\author{Saskia Grunau$$ and Valeria Kagramanova$$ \\ \\
$$Institut f\"ur Physik, Universit\"at Oldenburg,
D--26111 Oldenburg, Germany}

\maketitle

\begin{abstract}

We present the full set of analytical solutions of the geodesic equations of charged test particles in the
Reissner-Nordstr\"om space-time in terms of the Weierstra{\ss} $\wp$, $\sigma$ and $\zeta$ elliptic functions. 
Based on the study of the polynomials in the $\vartheta$ and $r$ equations we characterize the motion of test particles and discuss their properties. 
The motion of charged test particles in the
Reissner-Nordstr\"om space-time is compared with the motion of neutral test particles 
in the field of a gravitomagnetic monopole. 
Electrically or magnetically charged particles in the Reissner-Nordstr\"om space-time with magnetic or electric charges, respectively, move on cones 
similar to neutral test particles in the Taub-NUT space-times.

\end{abstract}

\section{Introduction}

The Reissner-Nordstr\"om space-time is a static, asymptotically flat solution
 of the Einstein-Maxwell equations in general relativity. 
It describes charged, non-rotating spherical black holes
or naked singularities. 
In general, the Reissner--Nordstr\"om space-time represents
a gravitating source which is both electrically and magnetically charged. 
It is a special case of the general family of electrovacuum space--times of Petrov type D found by Pleba\'{n}ski and Demia\'{n}ski~\cite{PlebanskiDemianski76} 
and reviewed by Griffiths and Podolsk\'{y}~\cite{GriffithsPodolsky06}. 
These space-times describe the gravitational fields of isolated massive objects,
e.g.~stars or black holes, 
and are in addition characterized by a NUT charge, angular momentum, acceleration and a cosmological constant. 
The higher dimensional generalizations of the Reissner-Nordstr\"om solution were found by Tangherlini~\cite{Tang63}.    

The physical properties of a space-time can be investigated by
studying the motion of test particles and light in this space-time. 
Test particles are a sensitive tool to investigate 
the properties of the fields generated by the massive body
because of the coupling of the parameters of the metric 
and the test particle in the equations of motion. 
The geodesic equations can be solved either analytically or numerically. 
A pioneer in finding analytical solutions was Hagihara~\cite{Hagihara31}. 
He integrated the geodesic equation of a test particle 
in the Schwarzschild gravitational field in terms of 
the Weierstrass $\wp$--function. 
Some special cases of the geodesics in the Reissner-Nordstr\"om space-time 
were studied in~\cite{Chandrasekhar83, Gack83}. 
General features of the motion of charged test particles 
in the Kerr-Newman space-time were considered in~\cite{Jiri89}. 
The influence of the gravitomagnetic monopole moment on the motion 
of test particles in Kerr-Newman-Taub-NUT space-time was 
studied in~\cite{Binietal03}. 

A breakthrough in the analytical integration of geodesics 
came with the papers of Hackmann and L\"ammerzahl~\cite{HackmannLaemmerzahl08_PRL,HackmannLaemmerzahl08_PRD} where geodesics in a 4-dimensional 
Schwarzschild-de Sitter space-time were integrated analytically 
in terms of the hyperelliptic $\theta$ and $\sigma$ functions. 
The developed method is based on the Jacobi inversion problem 
restricted to the $\theta$-divisor. 
Also geodesics in the Reissner-Nordstr\"om-de Sitter space-time 
in 4 dimensions as well as in some higher dimensional 
Schwarzschild, Schwarzschild--(anti)de Sitter, 
Reissner--Nordstr\"om and Reissner--Nordstr\"om--(anti)de Sitter space-times 
were integrated by this mathematical method~\cite{HD2008}. 
The elliptic and hyperelliptic functions were subsequently used 
to obtain the solutions of the geodesic equations 
in the axially symmetric Taub-NUT~\cite{taubnut2010} 
and Kerr-de Sitter~\cite{KerrMeth} space-times. 
In these papers the types of orbits are classified 
and the orbits of test particles are extensively studied. 
Furthermore, the analytical solution of the geodesic equations 
in the general Pleba\'{n}ski-Demia\'{n}ski space-time in 4 dimensions 
were obtained
in terms of the hyperelliptic Kleinian $\sigma$ function~\cite{PleDemMeth}.   

The Reissner-Nordstr\"om space-time as well as the Schwarzschild space-time 
have singularities at their origin. 
Whereas in the Schwarzschild space-time
particles are bound to end up in the central singularity,
this does not hold for particles which
cross the event horizon in the charged space-time. 
Here the potential barrier due to the charge does not allow the singularity 
to swallow the particles. 
Instead, a (test) particle will leave the vicinity of the singularity
again. After traversing the Cauchy horizon and the event horizon
of the  Reissner-Nordstr\"om black hole,
the particle will emerge in another 
universe~\cite{Chandrasekhar83,Straumann,HD2008}. 
Moreover, an orbit in such a black hole space-time
can be analytically continued into 
an infinite sequence of further patches of the space-time. 
This feature of the Reissner-Nordstr\"om solution
is seen clearly in its Carter-Penrose diagram. 

The Taub-NUT space-time which is characterized by a gravitomagnetic charge 
possesses also two horizons. 
A Kruskal-like analytic extension~\cite{MiKruGo71} 
of the Taub-NUT space-time would then likewise allow test particles 
to move into other worlds.
However, such an extension
comes at the prize that the periodic identification of the time
coordinate, suggested by Misner to remove the singularity
from the symmetry axis, is no longer possible~\cite{MiKruGo71}.
Consequently, the space-time is geodesically incomplete,
since orbits end either at the singular axis
(in the Bonnor-Manko-Ruiz interpretation)
or at one of the horizons
(in the Misner-Taub interpretation).
For a recent discussion see~\cite{taubnut2010}.

But the presence of two horizons and the associated many-world orbits
is not the only similarity between the Reissner-Nordstr\"om space-time
with electromagnetic charge and the Taub-NUT space-time
with gravitomagnetic charge. 
Also the motion of test particles has common properties
in both types of space-times.
In this paper we study the motion of electrically and/or magnetically 
charged particles in the general 
Reissner-Nordstr\"om space-time. 
As in the Taub-NUT spacetime where a test particle moves on a cone, 
the motion of a charged test particle in the 
general Reissner-Nordstr\"om space-time 
proceeds on a cone.
The condition for the motion on a cone is 
that the products of the dual charges should not be equal
i.e., $Gq \ne Qg$
(where $Q$ and $G$ are the electric and magnetic
charges of the gravitational source 
and $q$ and $g$ those of the test particle,
see Section~\ref{geodeq} for details).
Otherwise the motion would be planar.

In the following section we present the equations ot motion.
We classify the orbits of charged test particles 
in Section~\ref{theta-r-features}.
We analyze their motion by studying the influence of the parameters 
of the metric and of the test particles.
In particular, we consider the dependence
on the mass, on the angular momentum, on the charges and 
on the separation (Carter) constant. 
In Section~\ref{sec:solutions}
we solve analytically the $r$, $\vartheta$, $\varphi$ and $t$ 
equations of motion, and we give 
the solutions of the radial and time equations 
in terms of the Weierstrass elliptic $\wp$, $\sigma$ and $\zeta$ functions.  
We give details on the elliptic integrals in the Appendices.

\section{The geodesic equation}~\label{geodeq}

The Reissner-Nordstr\"om solution of the Einstein-Maxwell field equations is described by the metric~\cite{GriffithsPodolsky06} 
\begin{equation}
ds^2 = \frac{\Dr}{r^2} dt^2 - \frac{r^2}{\Dr}dr^2 - r^2\left( d\vartheta ^2 + \sin^2\vartheta  d\varphi ^2 \right) \, , \label{metrikNUTdeSitter}
\end{equation}
where $\Dr= r^2 - 2 M r + {Q}^2 + {G}^2 $. 
Here $M$ is the mass of the solution, 
${Q}$ and ${G}$ are the electric and magnetic charges. 
The singularity is located at $r=0$. 
For $0<Q^2+G^2<M^2$,
there are two horizons, defined by $\Delta_r = 0$, and given by
\begin{equation}
r_\pm = M \left( 1 \pm \sqrt{ M^2 - \left( Q^2 + G^2 \right)} \right) \, .
\end{equation}
Between the horizons the radial coordinate $r$ becomes timelike, 
and the time coordinate $t$ spacelike. 
When $Q^2+G^2 = M^2$, the horizons degenerate 
and the black hole is called extremal. 
For $Q^2+G^2 > M^2 $ the solution has a naked singularity.

The field strength $F_{\mu\nu}=A_{\nu, \mu} - A_{\mu, \nu}$ and the dual field strength $\check{F}_{\mu\nu}=\check{A}_{\nu, \mu} - \check{A}_{\mu, \nu}$ of the electromagnetic field are induced by the non-vanishing components of the vector potentials $A_\mu$ and ${\check{A}_\mu}$
\begin{eqnarray}
A_t=\frac{Q}{r} \ , \qquad A_\varphi=-G\cos\vartheta \ , \\
\check{A}_{t}=i\frac{G}{r} \ , \qquad \check{A}_{\varphi} = i Q \cos\vartheta \ .
\end{eqnarray}
The dual field strength is defined by the antisymmetric Levi-Civita symbol $\varepsilon^{\mu\nu\sigma\tau}$ as $\check{F}^{\mu\nu}=\frac{i}{2\sqrt{g^{\rm d}}}\varepsilon^{\mu\nu\sigma\tau}F_{\sigma\tau}$ with $g^{\rm d}=-\text{det}||g_{\mu\nu}||$.

The Hamilton-Jacobi equation for a particle 
with electric charge $q$ and magnetic charge $g$
\begin{equation}
2\frac{\partial S}{\partial\tau} = g^{ \mu\nu}\left(\frac{\partial S}{\partial x^\mu} - qA_\mu + ig\check{A}_{\mu} \right)\left( \frac{\partial S}{\partial x^\nu} - qA_\nu + ig\check{A}_{\nu} \right) \label{HJeq}
\end{equation}
has a solution in the form 
$S=\half\delta\tau - E t + L \varphi + S(r) + S(\vartheta)$. 
Here $\tau$ is an affine parameter along the geodesic, 
$\delta$ is a parameter which is equal to $0$ for a massless test particle 
and equal to $1$ for a test particle with non-zero mass. 
The constants $E$ and $L$ are the conserved energy and the angular momentum 
in the $z$ direction of a test particle, respectively. 
Because of the presence of the angle dependent 
terms $qA_\varphi$ and $g\check{A}_\varphi$ in~\eqref{HJeq} 
the problem of charged particle motion is axially symmetric, 
although the space-time itself as
described by the metric~\eqref{metrikNUTdeSitter} 
is spherically symmetric.

For convenience, we introduce dimensionless quantities ($r_{\rm S} = 2 M$)
\begin{equation}
\label{normpar}
\tr=\frac{r}{r_{\rm S}} \ , \,\, \tilde{t}=\frac{t}{r_{\rm S}} \ , \,\, \tilde{\tau}=\frac{\tau}{r_{\rm S}} \ , \,\,  \tilde{Q}=\frac{{Q}}{r_{\rm S}} \ , \,\, \tilde{G}=\frac{{G}}{r_{\rm S}} \ , \,\, \tilde{L}=\frac{L}{r_{\rm S}} \ .
\end{equation}

The Hamilton-Jacobi equation~\eqref{HJeq} separates and yields for each coordinate a corresponding differential equation
\begin{eqnarray}
\left(\frac{d\tr}{d\gamma}\right)^2 & = & R \label{eq-r-theta:1} \\
\left(\frac{d\vartheta }{d\gamma }\right)^2 & = & \Theta \,  \label{eq-r-theta:2} \\ 
\frac{d\varphi}{d\gamma} & =  & \frac{1}{\sin^2\vartheta}\left(\tL + \Dg \cos\vartheta  \right) \label{dvarphidgamma} \\
\frac{d \tilde{t}}{d\gamma} & = & \frac{\tr^4}{\tDr} \left( E + \frac{\De}{\tr} \right)  \, ,  \label{dtildetdgamma}
\end{eqnarray}
with the polynomial $R$ and the function $\Theta$
\begin{eqnarray}
R & = & \tr^4 \left( E + \frac{\De}{\tr} \right)^2-\tDr\left(\delta \tr^2 + k \right) \label{R_polynomial} \\
\Theta & = & {k}-\frac{1}{\sin^2\vartheta}\left(\tL + \Dg \cos\vartheta \right)^2   \label{Theta_polynomial} \, . 
\end{eqnarray}
The following notations were introduced: 
$\Dg=\tg q- \tq g$, $\De=\tq q+ \tg g$ and $\tDr=\tr^2 - \tr + \tq^2 + \tg^2$. 
Here $k=K+\tilde{L}^2$, 
where $K$ is known as Carter constant or separation constant. 
For zero charges the equations of motion above reduce to the Schwarzschild case.
We also used the Mino time $\gamma$ as 
$\tr^2 d\gamma = d\tilde{\tau}$ \cite{Mino03}.
Note, that for $\Dg = 0$ the situation simplifies,
and the motion takes place in a plane.

\section{Complete classification of geodesics}\label{theta-r-features}

Consider the Hamilton-Jacobi equations~\eqref{eq-r-theta:1}-\eqref{dtildetdgamma}. The properties of the orbits are given by the polynomial $R$~\eqref{R_polynomial} and the function $\Theta$~\eqref{Theta_polynomial}. The constants of motion (energy, angular momentum and separation constant) as well as the parameters of the metric and the charges of the test particle characterize these polynomials and, as a consequence, the types of orbits. In this section we discuss the charged motion in the Reissner-Nordstr\"om space--times in terms of the properties of the underlying polynomial $R$ and the function $\Theta$.

\subsection{The $\vartheta$--motion}\label{subsec:theta-pot}

Since $\vartheta$ is a polar angle, it can take only real values. Equation~\eqref{eq-r-theta:2} has real solutions if the requirement $\Theta\geq 0$ is fulfilled. 
This implies $ {k}  \geq 0 $.
With the new variable $\xi =  \cos\vartheta$, Eq.~(\ref{eq-r-theta:2}) 
turns into the equation 
\begin{equation}
\left(\frac{d\xi}{d\gamma}\right)^2 = \Theta_\xi \quad \text{with} \quad \Theta_\xi =  a \xi^2 + b \xi + c \, , \label{xieom}
\end{equation}
with a simple polynomial of second order on the right hand side, where $a = - ({k} + {\Dgs})$, $b = - 2 \tilde{L} {\Dg}$, and $c = {k} - \tilde{L}^2$.
 From $ {k} \geq 0$ follows $a < 0$. 
The zeros of $\Theta_\xi$ define the angles of two cones which confine the motion of the test particles 
(a similar feature appears in Taub-NUT~\cite{taubnut2010} and Kerr space--times~\cite{KerrMeth}). Moreover, every trajectory is not only constrained by these cones but lies itself on a cone in 3--space~\cite{monocone1,monocone2}. If $\Dg$ in the equation~\eqref{dvarphidgamma} and in the polynomial~\eqref{Theta_polynomial} vanishes,
then the motion lies on a plane 
(e.g., the motion of only electrically charged or neutral particles 
reduces to a plane in a Reissner-Nordstr\"om space-time 
with only electric charge). 
In the space-time of a gravitomagnetic monopole the trajectories of 
test particles similarly lie on cones~\cite{MisnerTaub69,ZimSha89,BellZonos98}. 

The discriminant $D = b^2 - 4 a c$ of the polynomial $\Theta_\xi$ 
can be written as $D = 4 {k} {\kappa}$ with 
${\kappa} =  {k} + \Dgs -\tilde{L}^2$. 
The existence of real zeros of $\Theta_\xi$ requires $D \geq 0$.
This implies that both ${k}$ and ${\kappa}$ should be non-negative
\begin{equation} 
\begin{array}{rl}   {k} & \geq 0 \\
                   {\kappa} = {k} -\tilde{L}^2 + \Dgs  & \geq  0  \ . \end{array} \  \label{theta-cond-lamu} 
\end{equation}
These are conditions on the parameters $\tilde{L}$ and ${k}$ for 
given values of $\tq$, $\tg$, $q$ and $g$. 
As long as ${k} -\tilde{L}^2$ is positive there are no constraints on $\tilde{L}$. 
When ${k} -\tilde{L}^2$ becomes negative the inequalities 
\eqref{theta-cond-lamu} imply a lower limit 
for the angular momentum given by $\tilde L_{\rm min} = \pm\sqrt{\Dgs+{k}}$. 

One can show that
\begin{equation}
\Theta = {k}-\frac{1}{\sin^2\vartheta}\left(\tL + \cos\vartheta \Dg \right)^2 = {\kappa}-\frac{1}{\sin^2\vartheta}\left(\tL\cos\vartheta + \Dg \right)^2 \, \label{Theta_c1c2} .
\end{equation}
The zeros of $\Theta_\xi$ are given by
\begin{equation}
\xi_{1,2}= \frac{\tilde{L} \Dg \pm \sqrt{{k}{\kappa}}}{-({k} + \Dgs)} \ . \label{OpeningAngles}
\end{equation}
The conditions~\eqref{theta-cond-lamu} ensure the compatibility of $\xi \in [-1, 1]$ with $\Theta_\xi \geq 0$.

The function $\Theta_\xi$ describes a parabola with the maximum at $\displaystyle{\left(-\frac{\tilde{L}\Dg}{{k}+\Dgs}, \frac{{k}{\kappa}}{{k}+\Dgs}\right)}$. 
For non-vanishing $\tilde L$ and $\Dg$ the maximum of the parabola is no longer located at $\xi = 0$ or, equivalently, the zeros are no longer symmetric with respect to $\xi = 0$. Only for vanishing $\tilde L$ or $\Dg$ both cones are symmetric with respect to the equatorial plane. 

The $\vartheta$--motion can be classified according to
the sign of ${k} - \tilde{L}^2$:
\begin{enumerate}  
\item If ${k} - \tilde{L}^2 < 0$ then $\Theta_\xi$ has 2 positive zeros for $\tilde{L}\Dg < 0$ and $\vartheta  \in  (0, \pi/2)$, so that the particle moves above the equatorial plane without crossing it. If $\tilde{L}\Dg > 0$ then $\vartheta  \in  (\pi/2, \pi)$.

\item If ${k} - \tilde{L}^2 = 0$ then $\Theta_\xi$ has two zeros: $\xi_1=0$ and $\xi_2=-\frac{2 \tilde{L} \Dg}{\tilde{L}^2 + \Dg^2}$. If $\tilde{L} \Dg < 0$ then $\xi \in [0, 1)$ and $\vartheta \in (0, \frac{\pi}{2}$]. If $\tilde{L} \Dg > 0$ then $\xi \in (-1, 0]$ and $\vartheta \in [\frac{\pi}{2}, \pi)$. If $\tilde{L} =  -\Dg$ 
then the $\vartheta$--motion fills the whole upper hemisphere $\vartheta \in [0, \frac{\pi}{2}]$. The motion fills the whole lower hemisphere with $\vartheta \in [\frac{\pi}{2}, \pi]$ if $\tilde{L} = \Dg$.

\item If ${k} - \tilde{L}^2 > 0$ then $\Theta_\xi$ has a positive and a negative zero and  $\vartheta \in (0,\pi)$,  and the particle crosses the equatorial plane during its motion. 
\end{enumerate}

In general, the second term of the function $\Theta$ 
in \eqref{Theta_c1c2} diverges for $\vartheta \rightarrow 0, \pi$. 
However, if $\tilde{L} = - \Dg$ this term is regular 
for $\vartheta = 0$ and if $\tilde{L} = \Dg$ it is regular for $\vartheta = \pi$. If $\tilde{L} = \pm \Dg$, then ${k} = {\kappa}$. 
The regularity of $\Theta$ in these cases can be seen from
\begin{equation}
\Theta = {\kappa} - \tilde{L}^2 \frac{(1 \mp \cos\vartheta)^2}{\sin^2\vartheta} \, , 
\end{equation}
by application of L'H\^{o}pital's rule. 
If, furthermore, $\tilde{L}^2 = 0$ then $\Theta = k + \Dgs - \tilde{L}^2$ which is independent of $\vartheta$. 

In the special cases when one of the constants ${k}$ or $\kappa$ or both vanish, 
$\Theta_\xi$ has a double root which is the only possible value for 
$\vartheta$. One can distinguish three cases:
\begin{enumerate}
\item If ${k} = 0$ and ${\kappa} > 0$, then $\xi = - \frac{ \tilde{L}}{\Dg}$ for $ \tilde{L}^2 < \Dgs $.
\item If ${\kappa} = 0$ and ${k} > 0$ then $\xi = - \frac{\Dg}{ \tilde{L}}$ for $ \tilde{L}^2 > \Dgs$.
\item If ${k} = {\kappa} = 0$ then $\xi = \pm 1$ implying that 
$\vartheta = 0$ or $\vartheta = \pi$ are possible. 
In this case $\tilde{L} = \mp \Dg$ (as discussed above).  
\end{enumerate} 
This means that during a test particle's motion the coordinate $\vartheta$ is constant and the trajectory lies on a cone around the $\vartheta=0, \pi$--axis with the opening angle $\arccos\xi$. In this case we immediately obtain from \eqref{dvarphidgamma} that $\varphi(\gamma) = {\cal C} (\gamma - \gamma_{\rm in})$ with a constant 
${\cal C} = \frac{\tilde{L} + \xi \Dg}{1-\xi^2}$ which in case 1 is ${\cal C} = 0$ and in case 2 is ${\cal C} = \tilde{L}$. 

Thus, the non-vanishing of ${k}$ and $\kappa$ indicates that the motion of the particle is not symmetric with respect to the $\vartheta=0, \pi$--axis. Therefore, these two constants may be regarded to play the role of a generalized Carter constant which appears, e.g., in the motion of particles in NUT and Kerr space--times.

\subsection{The $\tr$--motion}\label{subsec:r-motion}

\subsubsection{Possible types of orbits}

In the following we consider the motion of charged particles
in the regular Reissner-Nordstr\"om black hole space-time,
possessing two non-degenerate horizons $r_\pm$.
Before discussing the $\tr$--motion in detail,
we introduce a list of all possible orbits:
\begin{enumerate}\itemsep=-2pt
	\item {\it Escape orbits} (EO) with ranges: $(r_1, \infty)$ with $r_1 > r_+$. These escape orbits do not cross the horizons. 
	\item {\it Two-world escape orbits} (TEO) with ranges $(r_1, \infty)$ with $r_1 < r_-$
	\item {\it Periodic bound orbits} (BO) with range $\tr \in (r_1, r_2)$ with $r_1 < r_2$ and
\begin{enumerate}
	\item either $r_1, r_2  > r_+$ or 
	\item $r_1, r_2 < r_-$. 
\end{enumerate}
	\item {\it Many-world periodic bound orbits} (MBO) with range $\tr \in (r_1, r_2)$ where $r_1 < r_-$ and $r_2 > r_+$.  
\end{enumerate}
Here we used the notation for the orbits introduced in~\cite{HD2008} 
for the regular Reissner-Nordstr\"om space-time. 

\subsubsection{The radial motion}\label{radmotion}

\begin{figure}[th!]
\begin{center}
\subfigure[][$\tq=0.3, \tg=0.1, q=0.1, g=0$]{\label{nutn005k0}\includegraphics[width=6.9cm]{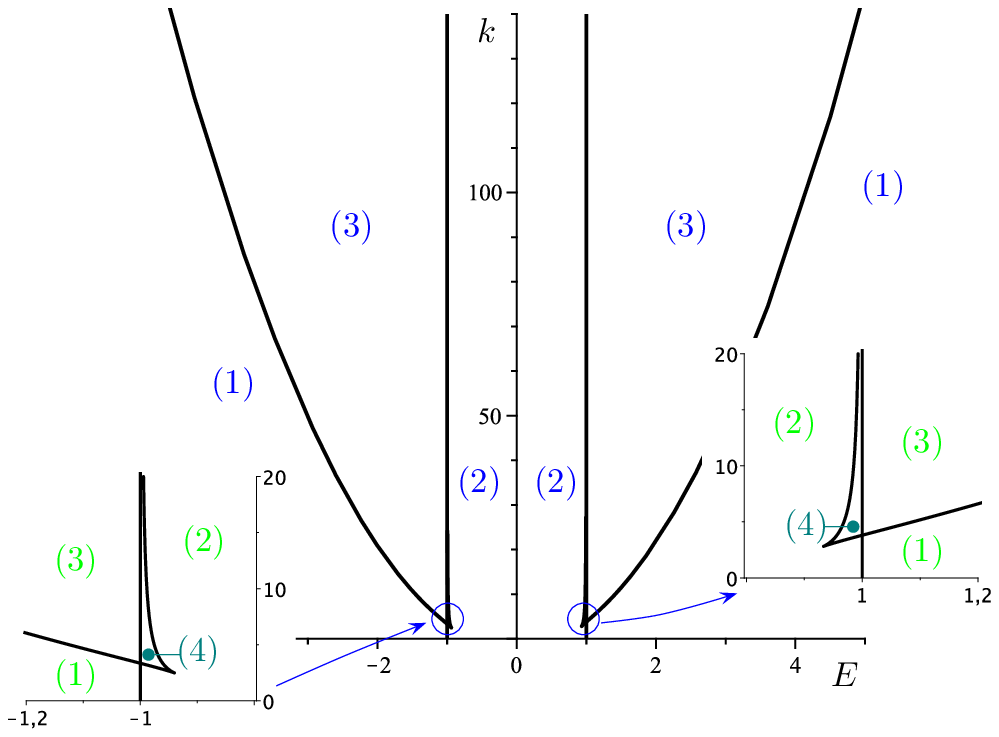}}
\subfigure[][$\tq=0.3, \tg=0.1, q=2.5, g=0$]{\label{nutn05k0}\includegraphics[width=6.9cm]{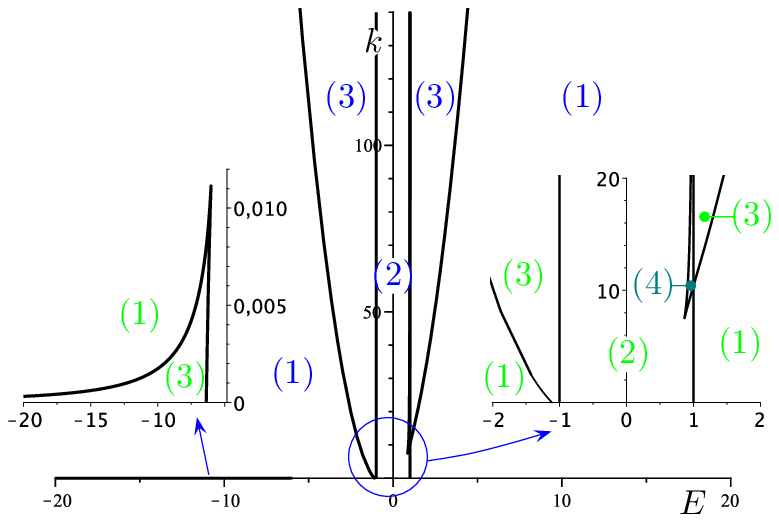}}
\subfigure[][$\tq=0.3, \tg=0.1, q=10, g=0$]{\label{nutn2k0}\includegraphics[width=6.9cm]{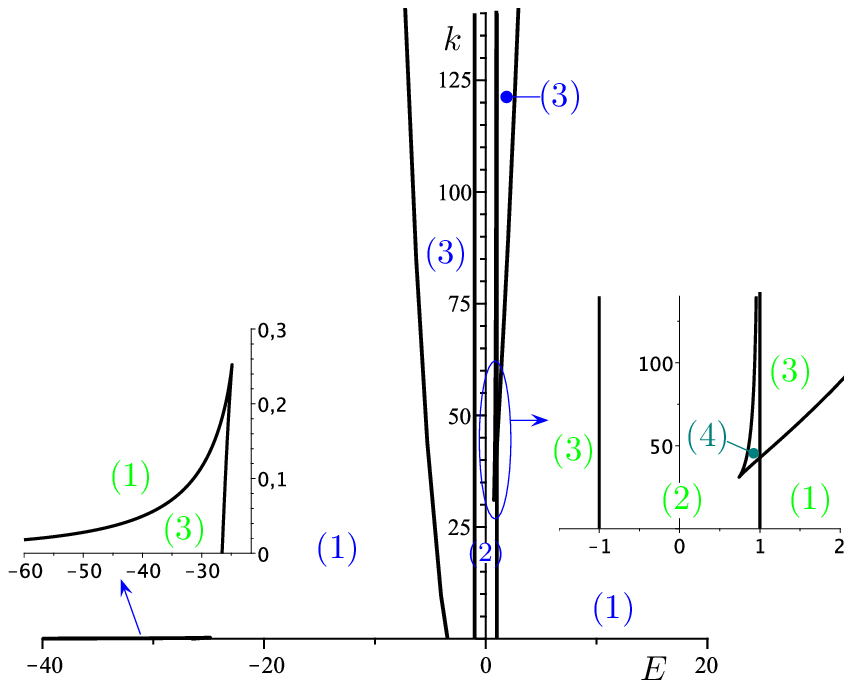}}
\subfigure[][$\tq=0.4, \tg=0.2, q=-3, g=0.1$]{\label{kEq04g02q-3}\includegraphics[width=6.9cm]{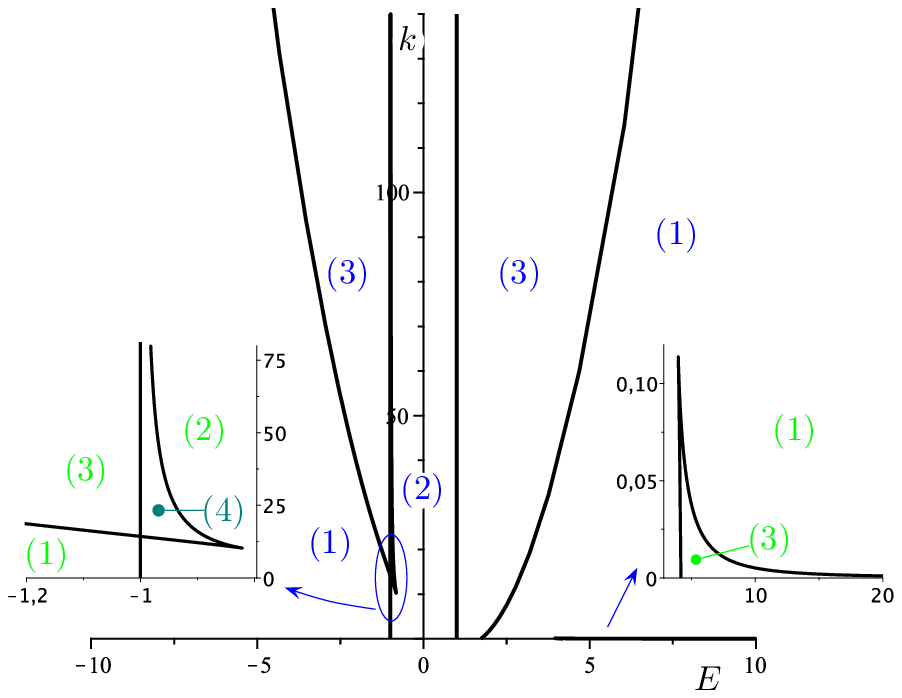}}
\end{center}
\caption{Parametric ${k}$-$E$ digrams showing the location of the 
regions $(1)$--$(4)$, which reflect the number of zeros of the polynomial $R$ 
in Eq.~\eqref{R_polynomial}. 
Each region contains a set of orbits peculiar to it which are described 
in Table~\ref{TypesOfOrbits1} and in the text (Section~\ref{radmotion}). 
Here the influence of the variation of the electric charge $q$ 
of a test particle on the zeros of $R$ is presented. 
Increasing the positive value of the electric charge 
in plots~\subref{nutn005k0}-\subref{nutn2k0} one observes, 
that the region $(4)$ with four positive zeros for negative $E$ 
(left side) disappears slowly. 
When the sign of $q$ is changed (along with the sign of $g$)
the right and left side of the ${k}$-$E$ diagrams are mirrored
(compare plot~\subref{kEq04g02q-3} with the rest). 
One finds similar ${k}$-$E$ diagrams by varying the value of the magnetic charge $g$. 
Because of the inequalities~\eqref{theta-cond-lamu} the constant ${k}=K+\tilde{L}^2$ is positive.  \label{nut_LE-diagrams}}
\end{figure}

The right hand side of the differential equation~\eqref{eq-r-theta:1} has the form $R=\sum^4_{i=0}{b_i \tr^i}$ with the coefficients 
\begin{eqnarray}
b_4 & = & {E}^2 - \delta  \\ 
b_3 & = & \delta + 2 E \De \label{Rcoefficients} \\ 
b_2 & = & - ( {k} - \Des + \delta(\tq{^2} + \tg^{2}) )   \\ 
b_1 & = & {k} \\ 
b_0 & = & -{k} (\tq{^2} + \tg^{2})  \, . 
\end{eqnarray} 
Let us now consider massive particles only, that is $\delta = 1$. 
In order to obtain real values for $\tr$ from \eqref{eq-r-theta:1} 
we have to require $R\geq0$. 
The regions for which $R \geq 0$ are bounded by the zeros of $R$. 
The number of zeros depends on the values of 
$E$, ${k}$, $\tq$, $\tg$, $q$ and $g$. 
Two conditions $R = 0$ and $\frac{dR}{d\tr} = 0$ define the double zeros of the polynomial $R$ and thereby the boundary between the regions where $R$ has $1$, $2$, $3$ or $4$ zeros. The parameter plots shown in Fig.~\ref{nut_LE-diagrams} are based on this. One has to additionally take care of the change of the sign of $E^2 - 1$ when $E$ crosses $E^2 = 1$. Then the sign of $R(\tilde r)$ for $\tilde r \rightarrow \pm \infty$ changes. Furthermore, the case $E^2 = 1$ requires additional attention: If the line $E^2 = 1$ is contained in a region with 3 or 1 zeros then on this portion of the line we have 3 or 1 zeros respectively. Taking all these features into account we obtain the ${k}$--$E$ diagrams of Fig.~\ref{nut_LE-diagrams}.

We define the effective potential from equation~\eqref{eq-r-theta:1} 
as the values of energy $E$ when 
\begin{equation}
0 = \left(\frac{d\tilde r}{d\gamma} \right)^2 =\tr^4 (E - V^+_{\rm eff})(E - V^-_{\rm eff}) \, , \label{turnpoints}
\end{equation}
thus,
\begin{equation}
V^\pm_{\rm eff} =  -\frac{\De}{\tr} \pm \frac{1}{\tr^2} \sqrt{ \tDr (\delta \tr^2 + {k}) } \label{Veff} \ .
\end{equation}
Condition~\eqref{turnpoints} then determines the turning points of an orbit.

\begin{figure}[htbp]
\centering
\subfigure[$\tq=0.4$ , $\tg=0.25$ , $q=0.05$ , $g=0.1$ , $k=4$: \newline
          Effective potential with four extrema.
          ]{\label{potentiale-1} 
\begin{overpic}[width=7cm]{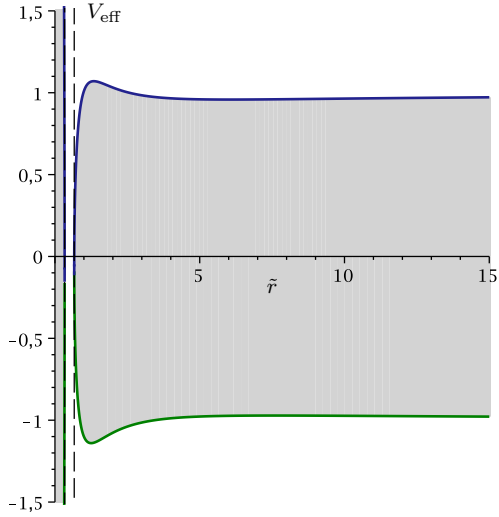}
\put(40,190){\footnotesize {$V_{\rm eff}$}}
\put(108,86){\footnotesize {$\tr$}}
\end{overpic}
}
\subfigure[Detail of figure (a):  1 to 4 turning points are possible.
 $V_{\rm eff}^\pm$ for orbits of type A, B, C, E from Table \ref{TypesOfOrbits1} showing the position of the orbit types BO, MBO, EO, TEO. The points indicate the turning points of the motion and the red horizontal dashed lines correspond to the values of the energy $E$.
          ]{\label{potentiale-2} 
\begin{overpic}[width=7cm]{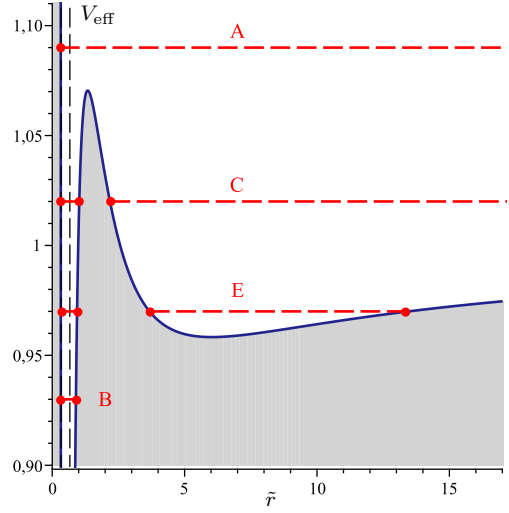}
\put(35,190){\footnotesize {$V_{\rm eff}$}}
\put(105,6){\footnotesize {$\tr$}}
\end{overpic}
}
\subfigure[$\tq=0.4$ , $\tg=0.2$ , $q=-3$ , $g=0.1$ , $k=1$.]	               
          {\label{potentiale-3}
          \begin{overpic}[width=7cm]{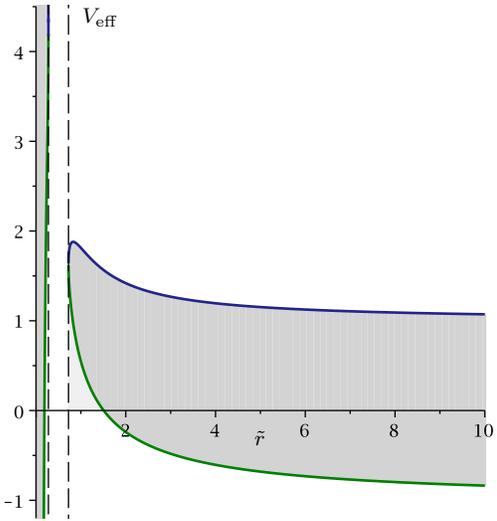}
          \put(40,190){\footnotesize {$V_{\rm eff}$}}
\put(105,30){\footnotesize {$\tr$}}
\end{overpic}
}
\subfigure[$\tq=0.4$ , $\tg=0.2$ , $q=3$ , $g=0.1$ , $k=1$.]
          {\label{potentiale-4} 
\begin{overpic}[width=7cm]{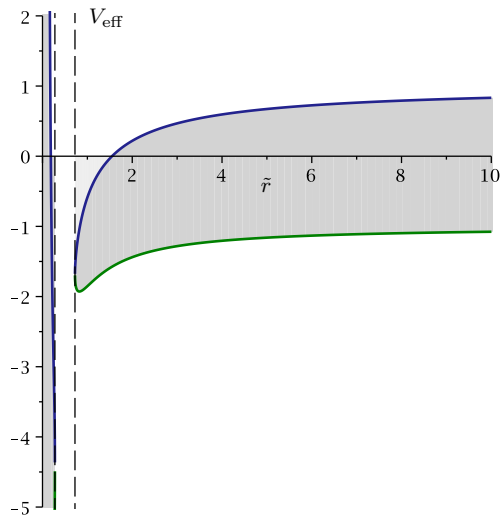}
\put(40,190){\footnotesize {$V_{\rm eff}$}}
\put(105,126){\footnotesize {$\tr$}}
\end{overpic}
}
          
\caption{Effective potential $V_{\rm eff}^\pm(\tr)$ for different values of $\tq$, $\tg$, $q$, $g$ and $k$. The blue line represents $V_{\rm eff}^+$ , the green line represents $V_{\rm eff}^-$. 
The two potential parts glue at the horizons--glue points--with the ordinate $V^{\pm}_{\rm}(r_\pm)=-\frac{\De}{r_\pm}$. 
The grey area marks a physically forbidden zone. 
The positions of the horizons are shown by vertical dashed lines. 
At infinity the effective potential tends to the limiting values 
$\lim_{\tr \to \infty} V_{\rm eff}^\pm = \pm 1$. 
Figures~\subref{potentiale-3} and~\subref{potentiale-4} 
are plotted for opposite electric charges of the test particle. 
The glue points of the $V^\pm_{\rm eff}$ are either positive (for $q<0$) 
or negative (for $q>0$). 
As can be seen from the ${k}$-$E$ diagram in Fig.~\ref{kEq04g02q-3} 
there are regions with $1$, $2$ or $3$ zeros for the 
potential~\subref{potentiale-3} with $k=1$.}
\label{potentiale}
\end{figure}

\begin{figure}[htbp]
\centering
\subfigure[$\tq=0.4$ , $\tg=0.25$ , $q=-4$ , $g=0.1$ , $k=0.2$]{\label{potentiale2-1}
\begin{overpic}[width=7cm]{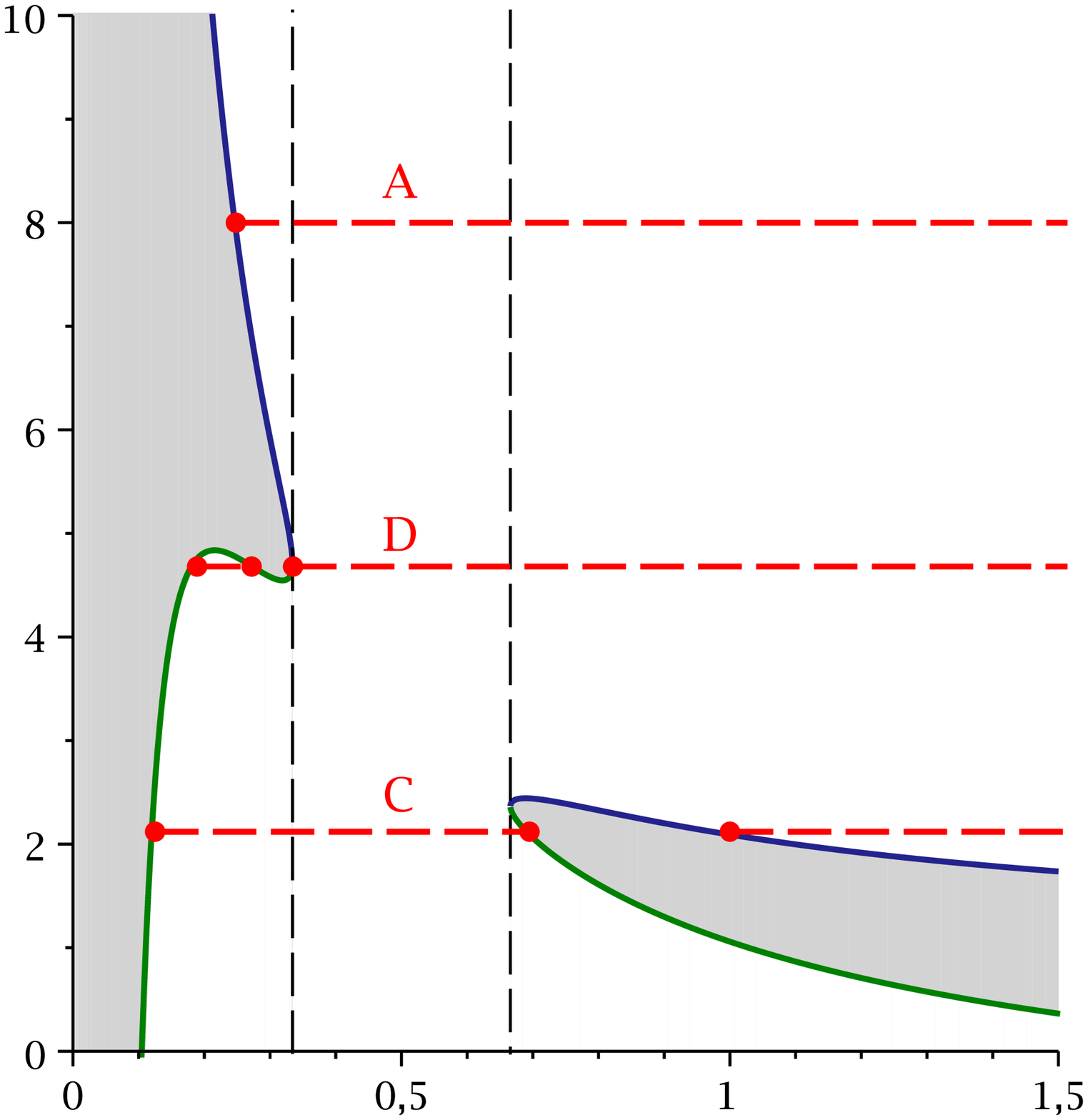}
\put(20,198){\footnotesize {$V_{\rm eff}$}}
\put(105,2){\footnotesize {$\tr$}}
\end{overpic}
}
\subfigure[$\tq=0.4$ , $\tg=0.25$ , $q=4$ , $g=0.1$ , $k=0.2$]{\label{potentiale2-2}
         \begin{overpic}[width=7cm]{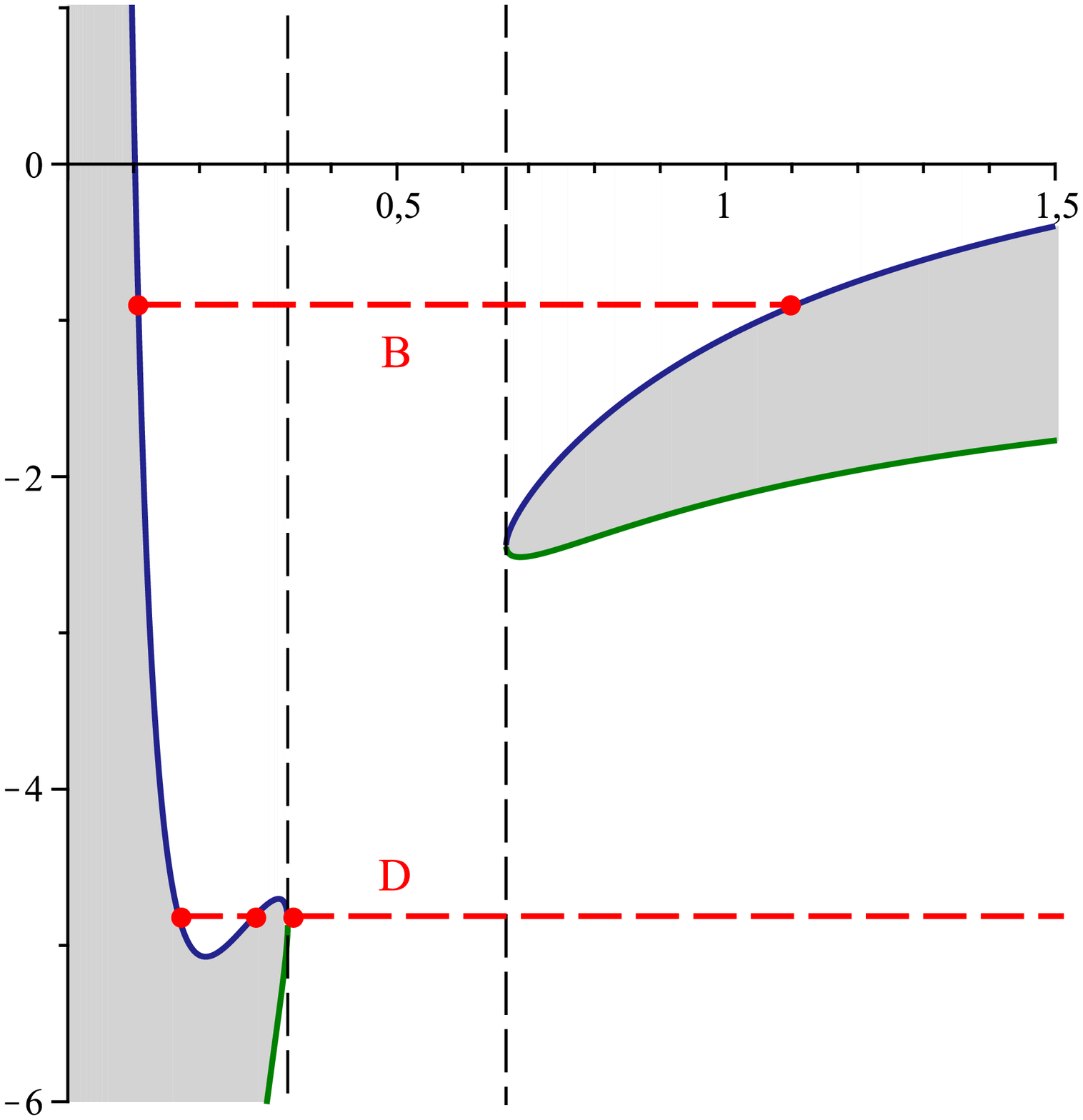}
\put(20,201){\footnotesize {$V_{\rm eff}$}}
\put(105,156){\footnotesize {$\tr$}}
\end{overpic}
}
\caption{Effective potential $V_{\rm eff}^\pm(\tr)$ 
for different values of $\tq$, $\tg$, $q$, $g$ and $k$. 
The grey area marks a physically forbidden zone. 
In this case up to 3 turning points exist for $\tr\le \tr_-$. 
Such an effect can be caused by a high charge of the test parcticle. 
In the plot~\subref{potentiale2-1} the orbits of type A, C, D from Table \ref{TypesOfOrbits1} are shown.}
\label{potentiale2}
\end{figure}

Examples for the effective potential are given 
in Figs.~\ref{potentiale}, \ref{potentiale2}. 
These figures combine 
the positive-root part $V^+_{\rm eff}$
and the negative-root part $V^-_{\rm eff}$.
At the horizons both parts $V^\pm_{\rm eff}$ coincide
(i.e.~at these points both parts $V^\pm_{\rm eff}$ are glued together),
since $\tDr$ vanishes here.
Consequently,
$V^\pm_{\rm eff}( r_\pm )=-\frac{\De}{r_\pm}$.
Clearly, $V^\pm_{\rm eff}( r_\pm )<0$ if $\De>0$ 
and vice versa (see e.g.~Fig.\ref{potentiale2}). 

We note, that in contrast to the
Schwarzschild case here the positive-root $V^+_{\rm eff}$
can allow for particles with negative energies. 
The region outside the horizon where negative energy particles may
sojourn has been termed `generalized ergospere',
since energy may be extracted
\cite{Christodoulou:1972kt,Denardo:1973}. 
While particle energies below the negative-root $V^-_{\rm eff}$
have no classical interpretation,
they are associated with antiparticles 
in the framework of quantum field theory
\cite{Deruelle:1974zy}.
The relation
$V^+_{\rm eff}(E,q,g) = - V^-_{\rm eff}(-E,-q,-g)$
shows, that the corresponding positive energy orbits are available
for particles with opposite charges (i.e., antiparticles).
A lower limit for the energies of particles is obtained
from the requirement that time should only run forward 
(see Eq.~(\ref{dtildetdgamma})),
$
E \ge - \frac{\De}{r}
$.

As we know from previous studies for neutral particles
(see e.g.~\cite{HD2008}),
the Reissner-Norstr\"om space-time possesses a potential barrier 
which prevents particles from falling into the singularity. 
The potential barrier which is defined 
by the smallest positive root $r_1$ of $R$ 
is located in the interval $0<r_1\leq r_-$. 

For increasing values of $\De$,
the influence of the charge of a test particle 
becomes noticeable through the term $-\frac{\De}{\tr}$.
(We recall, that the charges of the black hole
$\tq$ and $\tg$ are constrained by the 
naked singularity condition $\tq^2+\tg^2 > \frac{1}{4}$). 
Namely, for larger values of the test particle charge
the effective potential can form a small potential mound
(Fig.\ref{potentiale2-1}) or potential well (Fig.\ref{potentiale2-2}) 
in the interval $(0,r_-]$. 
This indicates that bound orbits may exist in this region. 
Such a feature is only present in the motion of charged particles.

Using the ${k}$--$E$ diagrams in Fig.~\ref{nut_LE-diagrams} as well as the above considerations we can give all possible combinations of zeros of $R$ and an interpretation in terms of specific types of orbits which are summarized in Table~\ref{TypesOfOrbits1}. 

The types of orbits related to the various parameters are given by:  
\begin{itemize}
\item Region (1): one positive zero. The orbit is a TEO with particles coming from $\tilde r = + \infty$. From the features of the effective potential (potential barrier) it is clear that the turning point can lie on the inner horizon (case $A_-$). 
In this case the energy of a test particle corresponds to the glue point of the potentials $V^\pm_{\rm eff}$: $E_{\rm A_-}=V^\pm_{\rm eff}\left(r_-\right)$. 
For $E^2=1$ the coefficient of the highest power in $R$, given by $E^2 - 1$ for massive test particles, vanishes. 
On the line $E^2=1$ within the region (1) there is one positive zero.
\item Region (2): two positive zeros. Here only MBOs are possible. 
In the special case when $\De=0$ and $E^2 = 0$, the turning points are lying on the horizons (case $B_\pm$). 
Also the cases $B_-$ and $B_+$ when one of the turning points 
lies on the horizons are possible. 
In this case the energy again corresponds to the glue points of the potentials 
$V^\pm_{\rm eff}$: $E_{\rm B_-}=V^\pm_{\rm eff}(r_-)$ 
(such an orbit can be found in 
e.g.~Fig.\ref{potentiale-1}) or $E_{\rm B_+}=V^\pm_{\rm eff}(r_+)$ 
(see e.g.~Fig.\ref{potentiale2-2}) respectively.  
\item Region (3): three positive zeros. 
\begin{itemize}
	\item Region $(3)_+$: Here MBOs and EOs with corresponding subcases $C_-$ and $C_+$ are possible. 
	\item Region $(3)_-$: For growing charges the term $-\frac{\De}{\tr}$ leads to a bound orbit with turning points behind the inner horizon. It is also possible that a turning point of the TEO lies on the inner horizon (case $D_-$).  
\end{itemize}
Here too for $E^2 = 1$ the term of highest power in $R$ vanishes giving three positive zeros for the $E^2=1$--line within the region (3). 

\item Region (4): four positive zeros. Here we find MBOs and BOs (planetary orbits). 
\end{itemize}

\begin{table}[t]
\begin{center}
\begin{tabular}{lcccl}\hline
type & region & +zeros & range of $\tilde r$ & orbit \\ \hline\hline
A & (1) & 1  & 
\begin{pspicture}(-2,-0.2)(3,0.2)
\psline[linewidth=0.5pt]{->}(-2,0)(3,0)
\psline[linewidth=0.5pt,doubleline=true](1.0,-0.2)(1.0,0.2)
\psline[linewidth=0.5pt,doubleline=true](-0.5,-0.2)(-0.5,0.2)
\psline[linewidth=1.2pt]{*-}(-1.1,0)(3,0)
\end{pspicture} 
 & TEO \\ 
${\rm A}_-$ & & & 
\begin{pspicture}(-2,-0.2)(3,0.2)
\psline[linewidth=0.5pt]{->}(-2,0)(3,0)
\psline[linewidth=0.5pt,doubleline=true](1.0,-0.2)(1.0,0.2)
\psline[linewidth=0.5pt,doubleline=true](-0.5,-0.2)(-0.5,0.2)
\psline[linewidth=1.2pt]{*-}(-0.5,0)(3,0)
\end{pspicture} 
 & ${\rm TEO}_-$ \\ \hline
B & $(2)$ & 2  & 
\begin{pspicture}(-2,-0.2)(3,0.2)
\psline[linewidth=0.5pt]{->}(-2,0)(3,0)
\psline[linewidth=0.5pt,doubleline=true](1.0,-0.2)(1.0,0.2)
\psline[linewidth=0.5pt,doubleline=true](-0.5,-0.2)(-0.5,0.2)
\psline[linewidth=1.2pt]{*-*}(-1.0,0)(1.5,0)
\end{pspicture} 
& MBO \\  
${\rm B}_+$ & &  & 
\begin{pspicture}(-2,-0.2)(3,0.2)
\psline[linewidth=0.5pt]{->}(-2,0)(3,0)
\psline[linewidth=0.5pt,doubleline=true](1.0,-0.2)(1.0,0.2)
\psline[linewidth=0.5pt,doubleline=true](-0.5,-0.2)(-0.5,0.2)
\psline[linewidth=1.2pt]{*-*}(-0.8,0)(1,0)
\end{pspicture} 
& ${\rm MBO}_+$ \\ 
${\rm B}_-$ & &  & 
\begin{pspicture}(-2,-0.2)(3,0.2)
\psline[linewidth=0.5pt]{->}(-2,0)(3,0)
\psline[linewidth=0.5pt,doubleline=true](1.0,-0.2)(1.0,0.2)
\psline[linewidth=0.5pt,doubleline=true](-0.5,-0.2)(-0.5,0.2)
\psline[linewidth=1.2pt]{*-*}(-0.5,0)(1.3,0)
\end{pspicture} 
& ${\rm MBO}_-$ \\ 
${\rm B}_\pm$ & &  & 
\begin{pspicture}(-2,-0.2)(3,0.2)
\psline[linewidth=0.5pt]{->}(-2,0)(3,0)
\psline[linewidth=0.5pt,doubleline=true](1.0,-0.2)(1.0,0.2)
\psline[linewidth=0.5pt,doubleline=true](-0.5,-0.2)(-0.5,0.2)
\psline[linewidth=1.2pt]{*-*}(-0.5,0)(1.0,0)
\end{pspicture} 
& ${\rm MBO}_\pm$ \\ \hline
C & $(3)_+$ & 3 & 
\begin{pspicture}(-2,-0.2)(3,0.2)
\psline[linewidth=0.5pt]{->}(-2,0)(3,0)
\psline[linewidth=0.5pt,doubleline=true](1.0,-0.2)(1.0,0.2)
\psline[linewidth=0.5pt,doubleline=true](-0.5,-0.2)(-0.5,0.2)
\psline[linewidth=1.2pt]{*-*}(-1.0,0)(1.5,0)
\psline[linewidth=1.2pt]{*-}(2.0,0)(3,0)
\end{pspicture} 
 & MBO, EO \\ 
 ${\rm C}_-$ & & & 
\begin{pspicture}(-2,-0.2)(3,0.2)
\psline[linewidth=0.5pt]{->}(-2,0)(3,0)
\psline[linewidth=0.5pt,doubleline=true](1.0,-0.2)(1.0,0.2)
\psline[linewidth=0.5pt,doubleline=true](-0.5,-0.2)(-0.5,0.2)
\psline[linewidth=1.2pt]{*-*}(-0.5,0)(1.5,0)
\psline[linewidth=1.2pt]{*-}(2.0,0)(3,0)
\end{pspicture} 
 & MBO, EO \\ 
 ${\rm C}_+$ &  &  & 
\begin{pspicture}(-2,-0.2)(3,0.2)
\psline[linewidth=0.5pt]{->}(-2,0)(3,0)
\psline[linewidth=0.5pt,doubleline=true](1.0,-0.2)(1.0,0.2)
\psline[linewidth=0.5pt,doubleline=true](-0.5,-0.2)(-0.5,0.2)
\psline[linewidth=1.2pt]{*-*}(-1.0,0)(1,0)
\psline[linewidth=1.2pt]{*-}(2.0,0)(3,0)
\end{pspicture} 
 & ${\rm MBO}_+$, EO \\
${\rm D}$ & $(3)_-$ & 3 &  
\begin{pspicture}(-2,-0.2)(3,0.2)
\psline[linewidth=0.5pt]{->}(-2,0)(3,0)
\psline[linewidth=0.5pt,doubleline=true](1.0,-0.2)(1.0,0.2)
\psline[linewidth=0.5pt,doubleline=true](-0.5,-0.2)(-0.5,0.2)
\psline[linewidth=1.2pt]{*-*}(-1.8,0)(-1.1,0)
\psline[linewidth=1.2pt]{*-}(-0.8,0)(3,0)
\end{pspicture} 
 & BO, TEO \\
${\rm D}_-$ &  &  & 
\begin{pspicture}(-2,-0.2)(3,0.2)
\psline[linewidth=0.5pt]{->}(-2,0)(3,0)
\psline[linewidth=0.5pt,doubleline=true](1.0,-0.2)(1.0,0.2)
\psline[linewidth=0.5pt,doubleline=true](-0.5,-0.2)(-0.5,0.2)
\psline[linewidth=1.2pt]{*-*}(-1.8,0)(-1.1,0)
\psline[linewidth=1.2pt]{*-}(-0.5,0)(3,0)
\end{pspicture} 
 & BO, TEO \\ \hline
E & $(4)$ & 4 &  
\begin{pspicture}(-2,-0.2)(3,0.2)
\psline[linewidth=0.5pt]{->}(-2,0)(3,0)
\psline[linewidth=0.5pt,doubleline=true](1.0,-0.2)(1.0,0.2)
\psline[linewidth=0.5pt,doubleline=true](-0.5,-0.2)(-0.5,0.2)
\psline[linewidth=1.2pt]{*-*}(-1.0,0)(1.5,0)
\psline[linewidth=1.2pt]{*-*}(2.0,0)(2.6,0)
\end{pspicture} 
 & MBO, BO  \\ \hline\hline
\end{tabular}
\caption{Types of polynomials and orbits of charged particles
 in the Reissner-Nordstr\"om space--time. 
The thick lines represent the range of the orbits. 
The turning points are shown by thick dots. 
The horizons are indicated by a vertical double line. 
In special cases the turning points lie on the horizons. 
Type $\rm D$ with a bound orbit behind the inner horizon 
is possible for charged particles with relatively large charges. 
\label{TypesOfOrbits1}}
\end{center}
\end{table}

\section{Solution of the geodesic equation}\label{sec:solutions}

Now we present the analytical solutions of the differential equations ~\eqref{eq-r-theta:1}--\eqref{dtildetdgamma}. 

\subsection{Solution of the $\vartheta$--equation}\label{vartheta-sol}

The solution of Eq.~\eqref{xieom} with $a<0$ and $D>0$ is given by the elementary function
\begin{equation}
\vartheta(\gamma )=\arccos\Bigl(\frac{1}{2a}\left(  \sqrt{D}\sin\left( \sqrt{-a}\gamma - \gamma^\vartheta_{\rm in}   \right) -b  \right) \Bigr) \ ,
\end{equation}
where $\gamma ^\vartheta_{\rm in}=\sqrt{-a}\gamma _{{\rm in}} - \arcsin \left( \frac{2a\xi_{{\rm in}}+b}{\sqrt{b^2-4ac}}\right)$ and $\gamma_{\rm in}$ is the initial value of $\gamma$.

\subsection{Solution of the $\tr$--equation}

For timelike geodesics the polynomial $R$ in~\eqref{R_polynomial} is of fourth order. 
A standard substitution $\tr=\pm\frac{1}{x} + \tr_R$, where $\tr_R$ is a zero of $R$, reduces \eqref{eq-r-theta:1} to a differential equation with a third order polynomial  $\left(\frac{dx}{d\gamma}\right)^2=R_3$, where $R_3=\sum^3_{i=0}{b_ix^i}$. A further substitution $x=\frac{1}{b_3}\left(4y-\frac{b_2}{3}\right)$ transforms that into the standard Weierstra{\ss} form
\begin{equation}
\left(\frac{dy}{d\gamma}\right)^2=4y^3-g_2y-g_3= P_3(y) \, ,  \label{P3}
\end{equation}
where 
\begin{equation}
g_2=\frac{b_2^2}{12} - \frac{b_1b_3}{4} \, , \qquad  g_3=\frac{b_1b_2b_3}{48} - \frac{b_0b_3^2}{16}-\frac{b_2^3}{216} \ .
\end{equation}
The differential equation \eqref{P3} is of elliptic type and is solved by the Weierstra{\ss} $\wp$--function~\cite{Markush}
\begin{equation}
y(\gamma) = \wp\left(\gamma - \gamma'_{\rm in}; g_2, g_3\right) \ , \label{soly}
\end{equation}
where $\gamma'_{\rm in}=\gamma_{\rm in}+\int^\infty_{y_{\rm in}}{\frac{dy}{\sqrt{4y^3-g_2y-g_3}}}$
with $y_{\rm in}=\pm\frac{b_3}{4}\left(\tr_{\rm in} - \tr_R\right)^{-1} + \frac{b_2}{12}$.
Then the solution of~\eqref{eq-r-theta:1} acquires the form
\begin{equation}
\tr=\pm \frac{b_3}{4 \wp\left(\gamma - \gamma'_{\rm in}; g_2, g_3\right) - \frac{b_2}{3}} + \tr_R \ . \label{solrNUTlight}
\end{equation}

\subsection{Solution of the $\varphi $--equation}

Eq.~\eqref{dvarphidgamma} can be simplified by using \eqref{eq-r-theta:2} 
and by performing the substitution $\xi = \cos\vartheta$ 
\begin{equation}
d\varphi = - \frac{d\xi}{\sqrt{\Theta_\xi}} \frac{\tL}{1 - \xi^2} - \frac{\xi d\xi}{\sqrt{\Theta_\xi}} \frac{\Dg}{1 - \xi^2} \label{phi_new} \, ,
\end{equation}
where $\Theta_\xi$ is given in \eqref{xieom}. This equation can be easily integrated and the solution for $a < 0$ and $D > 0$ is given by
\begin{equation}
\varphi(\gamma) = \half \Bigl(I_+ + I_-\Bigr)\Bigl|^{\xi(\gamma)}_{\xi_{{\rm in}}} + \varphi _{\rm in} \label{sol2phi} \, ,
\end{equation}
where
\begin{equation}
I_{\pm} = - \frac{ \tL \pm \Dg }{|\tL \pm \Dg|} \arcsin\frac{ k+\kappa-(\tL\pm\Dg)^2 \mp \left(k+\kappa+(\tL\pm\Dg)^2\right)\xi }{\sqrt{D}(\xi\mp 1)} \label{IpmEq}
\end{equation}

For the special case ${k}=\tL^2$ and $\tL = \pm \Dg$ the solution reduces to the simple form
\begin{equation}
\varphi (\gamma ) =  \half \frac{\tL}{|\tL|} 
\arcsin\frac{1 \pm 3 \xi}{ \xi \mp 1 }\Biggl|^{\xi(\gamma)}_{\xi_{{\rm in}}} + \varphi_{\rm in}   \label{sol2phi2_special} \ .
\end{equation}

\subsection{Solution of the $\tilde{t}$--equation}

Eq.~\eqref{dtildetdgamma} consists of two $\tr$-integrals:
\begin{eqnarray}
\tilde{t}-\tilde{t}_{\rm in}&=& {E} \int^{\tr(\gamma)}_{\tr_{\rm in}} \frac{\tr{^4}}{\tDr} \frac{d\tr}{\sqrt{R}} + {\De} \int^{\tr(\gamma)}_{\tr_{\rm in}} \frac{\tr{^3}}{\tDr} \frac{d\tr}{\sqrt{R}} =: I^1_{\tr}(\gamma) + I^2_{\tr}(\gamma) \ . \label{tint_NUT}
\end{eqnarray}

Consider $I^1_{\tr}$. 
The substitution $\tr=\pm \frac{b_3}{4y-\frac{b_2}{3}} + \tr_R$ 
reexpresses $I_{\tr}$ in terms of $y$:
\begin{equation}
I^1_{\tr}(\gamma)= E \int^y_{y_{\rm in}} \mp \frac{dy}{\sqrt{P_3(y)}} \frac{   \left(  \tr_{\rm R}\left(4y-\frac{b_2}{3}\right)\pm b_3  \right)^4}{\left(4y-\frac{b_2}{3}\right)^2\Delta_y} \ , \label{IryNUT}
\end{equation}
where $\Delta_y=\tDr(\tr_R)\left(4y-\frac{b_2}{3}\right)^2\pm\left(2 \tr_R -1 \right)b_3\left(4y - \frac{b_2}{3}\right) + b_3^2 = 16 \tDr(\tr_R) (y-p_1)(y-p_2)$. Here, $p_1$ and $p_2$ are two zeros of $\Delta_y$. 

We next apply a partial fractions decomposition upon Eq.~\eqref{IryNUT}
\begin{equation}
I^1_{\tr}(\gamma)= E \int^\gamma_{\gamma_{\rm in}} \mp \frac{dy}{\sqrt{P_3(y)}} \Biggl( K_0 + \sum^3_{j=1}\frac{K_j}{y-p_j} + \frac{K_4}{\left(y-p_3\right)^2} \Biggr) \ , \label{IryNUT2}
\end{equation}
where $p_3=\frac{b_2}{12}$ and $K_i$, $i=0,..,4$, 
are constants which arise from the partial fractions decomposition. These depend on the parameters of the metric 
and the test particle and on $\tr_{\rm R}$. 
After the substitution $y=\wp(v)$ with 
$\wp^\prime(v)=(-1)^\rho\sqrt{4 \wp^3(v)-g_2\wp(v)-g_3}$, 
where $\rho$ is either $0$ or $1$ depending 
on the sign of $\wp^\prime(v)$ in the considered interval 
and on the branch of the square root, 
Eq.~\eqref{IryNUT2} simplifies to
\begin{equation}
I^1_{\tr}(\gamma)= \mp E \int^v_{v_{\rm in}} (-1)^\rho \Biggl( K_0 + \sum^3_{j=1}\frac{K_j}{\wp(v)-p_j} + \frac{K_4}{\left(\wp(v)-p_3\right)^2} \Biggr)dv \ . \label{IryNUT3}
\end{equation}  
Here $v=v(\gamma)=\gamma-\gamma^\prime_{\rm in}$ and $v_{\rm in}=v(\gamma_{\rm in})$. 

The second integral $I^2_{\tr}$ in~\eqref{tint_NUT} 
can be reduced in a similar way to the necessary form, yielding:
\begin{equation}
I^2_{\tr}(\gamma)= \mp \De \int^v_{v_{\rm in}} (-1)^\rho \Biggl( H_0 + \sum^3_{j=1}\frac{H_j}{\wp(v)-p_j} \Biggr)dv \ , \label{IryNUT_I2}
\end{equation}  
where $H_i$, $i=0,..,3$, are constants which arise from the partial fractions decomposition.

The final solution takes the form (details can be found in the appendices \ref{elldiffIII} and \ref{elldiff2})
\begin{eqnarray}
I_{\tr}(\gamma) & = & \mp (-1)^\rho E \Biggl\{\left(K_0 + A_2K_4\right)(v-v_{\rm in}) +  \sum^2_{i=1}\Biggl[  \sum^3_{j=1} \frac{K_j}{\wp^\prime(v_{ji})}\Biggl( \zeta(v_{ji})(v-v_{\rm in}) + \log\frac{\sigma(v-v_{ji})}{\sigma(v_{\rm in}-v_{ji})}  \Biggr)  \nonumber \\
&& - \frac{K_4}{\left(\wp^\prime(v_{3i})\right)^2}\Biggl( \zeta(v-v_{3i}) - \zeta(v_{\rm in}-v_{3i}) + \frac{\wp^{\prime\prime}(v_{3i})}{\wp^\prime(v_{3i})}
 \log\frac{\sigma(v-v_{3i})}{\sigma(v_{\rm in}-v_{3i})} \Biggr) \Biggr] \Biggr\}  \nonumber \\
&& \mp (-1)^\rho \De \Biggl\{H_0(v-v_{\rm in}) +  \sum^2_{i=1}\Biggl[  \sum^3_{j=1} \frac{H_j}{\wp^\prime(v_{ji})}\Biggl( \zeta(v_{ji})(v-v_{\rm in}) + \log\frac{\sigma(v-v_{ji})}{\sigma(v_{\rm in}-v_{ji})}  \Biggr) \Biggr] \Biggr\} 
   \ , \label{IryNUT4}
\end{eqnarray} 
where $v_{ji}$ are the poles of the functions $(\wp(v)-p_j)^{-1}$ and $(\wp(v)-p_3)^{-2}$ in~\eqref{IryNUT2} such that $\wp(v_{j1})=p_j=\wp(v_{j2})$ since $\wp(v)$ is an even elliptic function of order two which assumes every value in the fundamental parallelogram with multiplicity two. Here $\zeta(v)$ is the Weierstra{\ss} zeta function and $\sigma(v)$ is the Weierstra{\ss} sigma function~\cite{Markush}; $\displaystyle{A_2=-\sum^{2}_{i=1}\left(  \frac{\wp(v_{3i})}{\left(\wp^\prime(v_{3i})\right)^2} + \frac{\wp^{\prime\prime}(v_{3i})\zeta(v_{3i})}{\left(\wp^\prime(v_{3i})\right)^3}  \right)}$ is a constant.

\subsection{The orbits}

\begin{figure}[htbp]
\centering
\subfigure[MBO.]{\includegraphics[width=6cm]{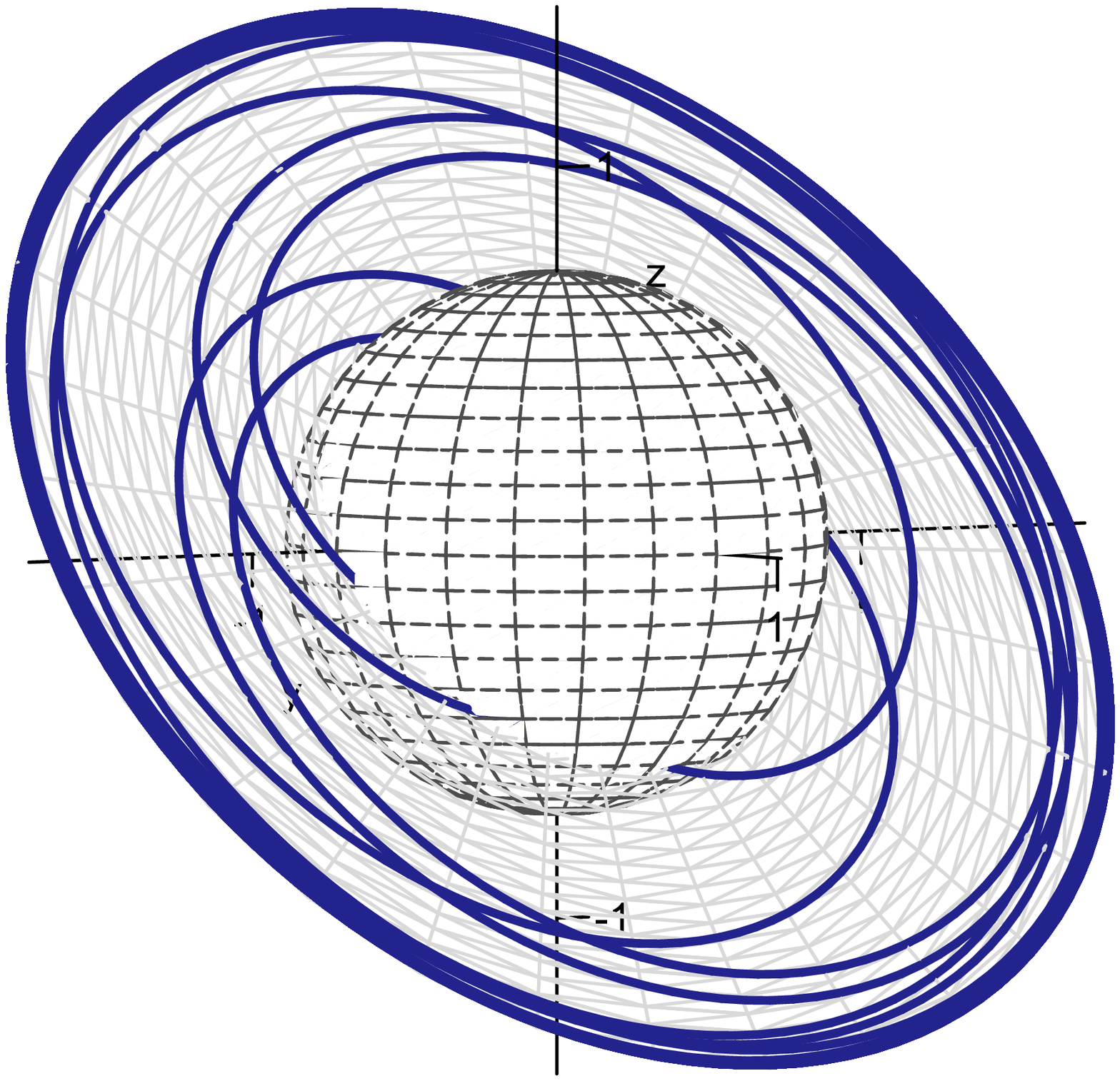}}
\subfigure[Projection onto the $x$-$z$-plane. 
The two circles denote the horizons $\tr_+$ and $\tr_-$. 
The potential barrier at $r<r_-$ prevents the particle 
from falling into the singularity.]
{\includegraphics[width=6cm]{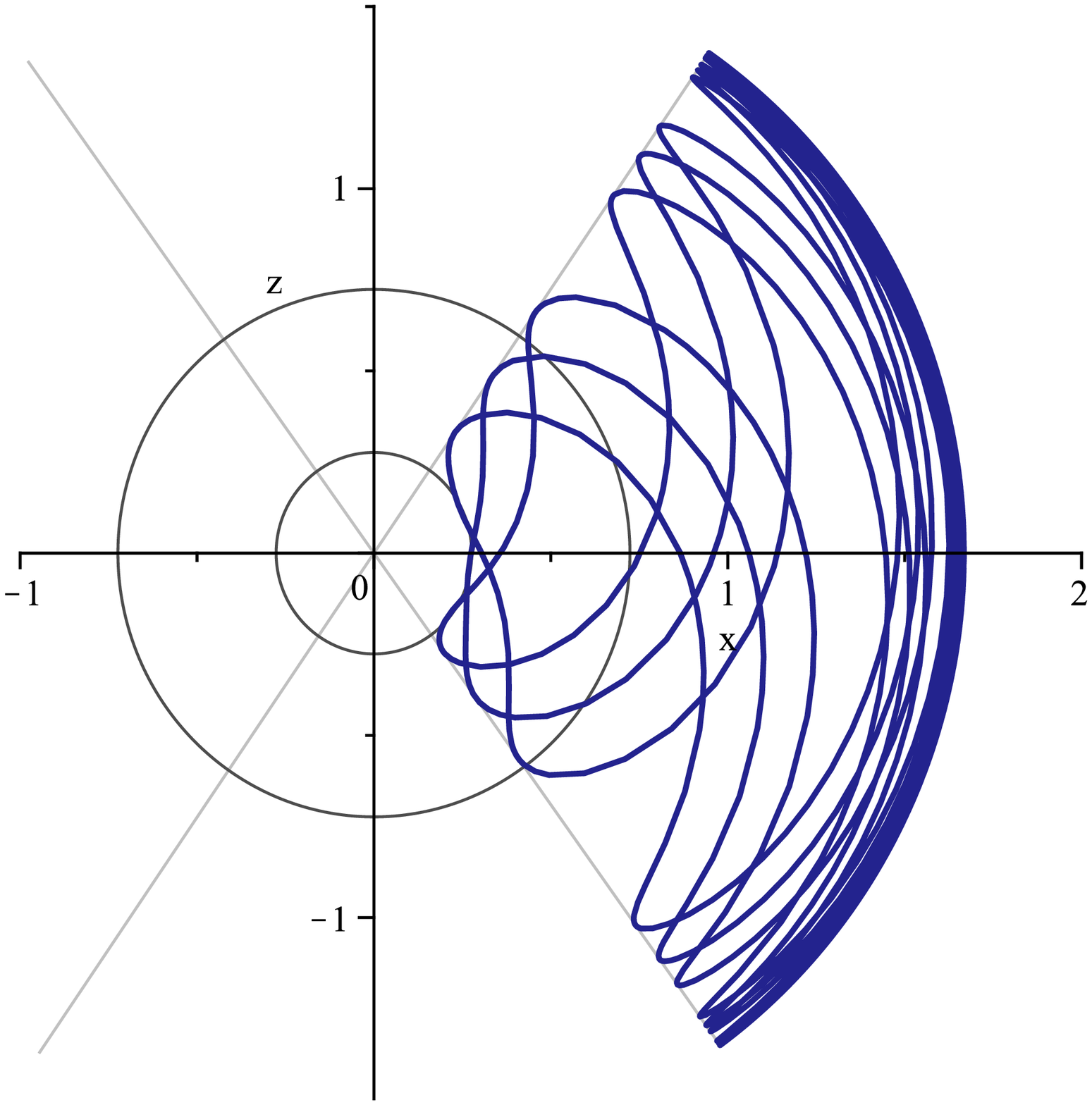}}
\subfigure[BO.]{\includegraphics[width=6cm]{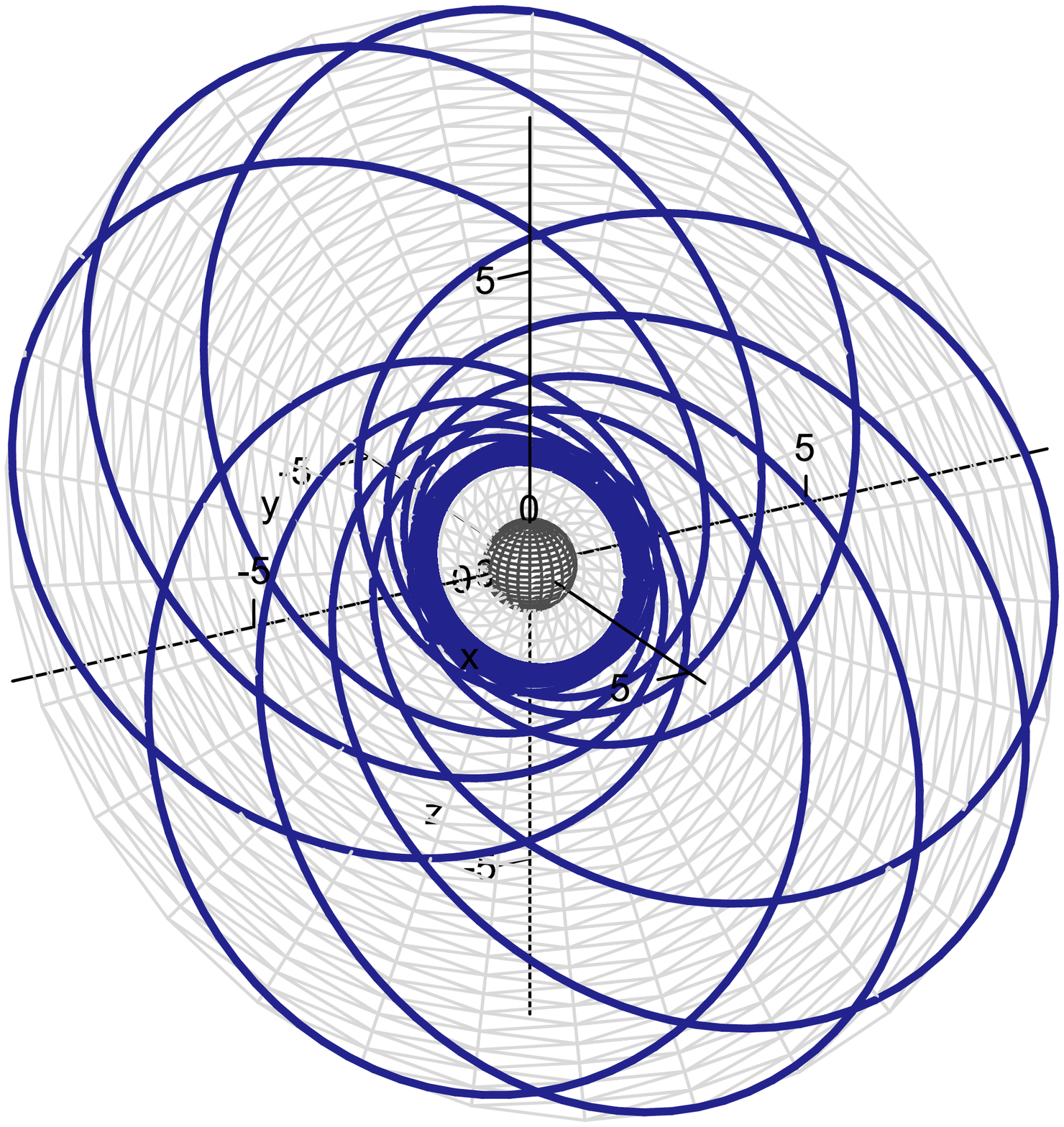}}
\subfigure[Projection onto the $x$-$z$-plane. The two circles denote $\tr_+$ and $\tr_-$.]{\includegraphics[width=6cm]{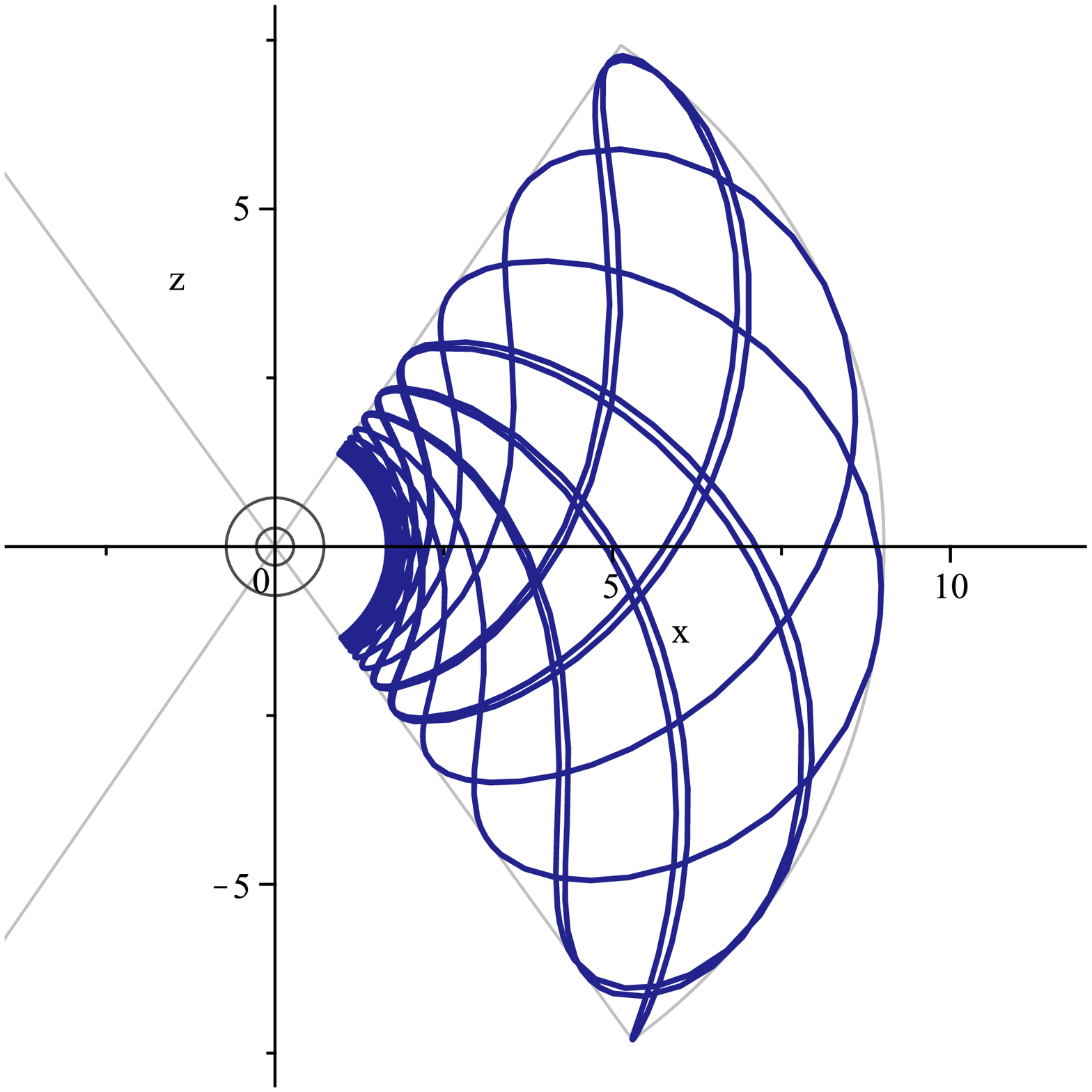}}
\caption{MBO and BO with parameters $\tq =0.4$, $\tg =0.2$, $q=0.1$, $g=0.1$, $k=3$, $E=0.9548$, $\tL=1.0$. The sphere in the 3d plots shows the horizon $\tr_+$. Orbits lie on cones with large opening angle $\Delta\vartheta\rightarrow\pi$.}
\label{MBO-1}
\end{figure}

\begin{figure}[htbp]
\centering
\subfigure[TEO. The sphere shows the horizon $\tr_+$.]
{\includegraphics[width=6cm]{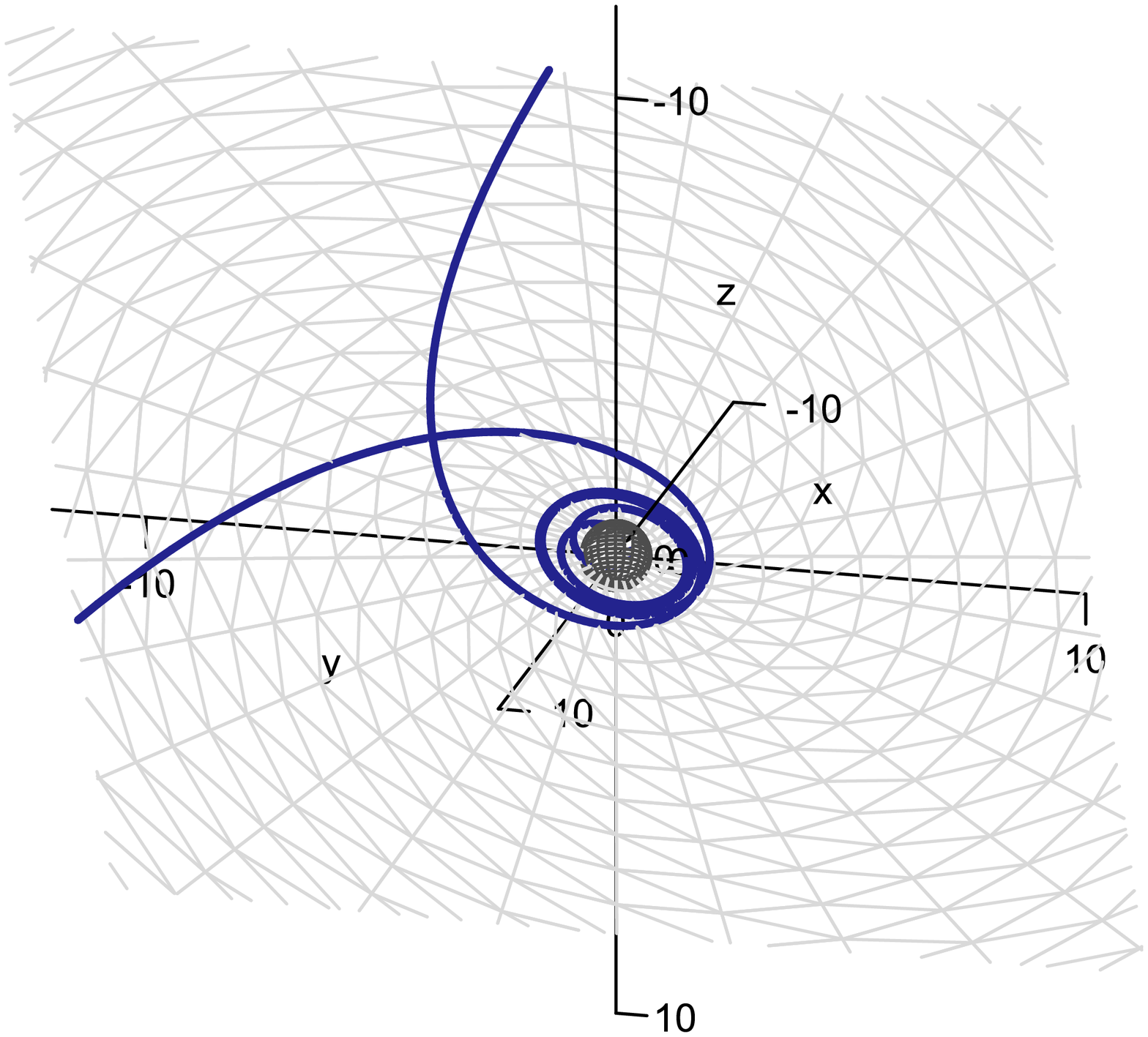}}
\subfigure[Projection onto the $x$-$z$-plane. The two circles denote $\tr_+$ and $\tr_-$.]{\label{TEO-1-2}\includegraphics[width=6cm]{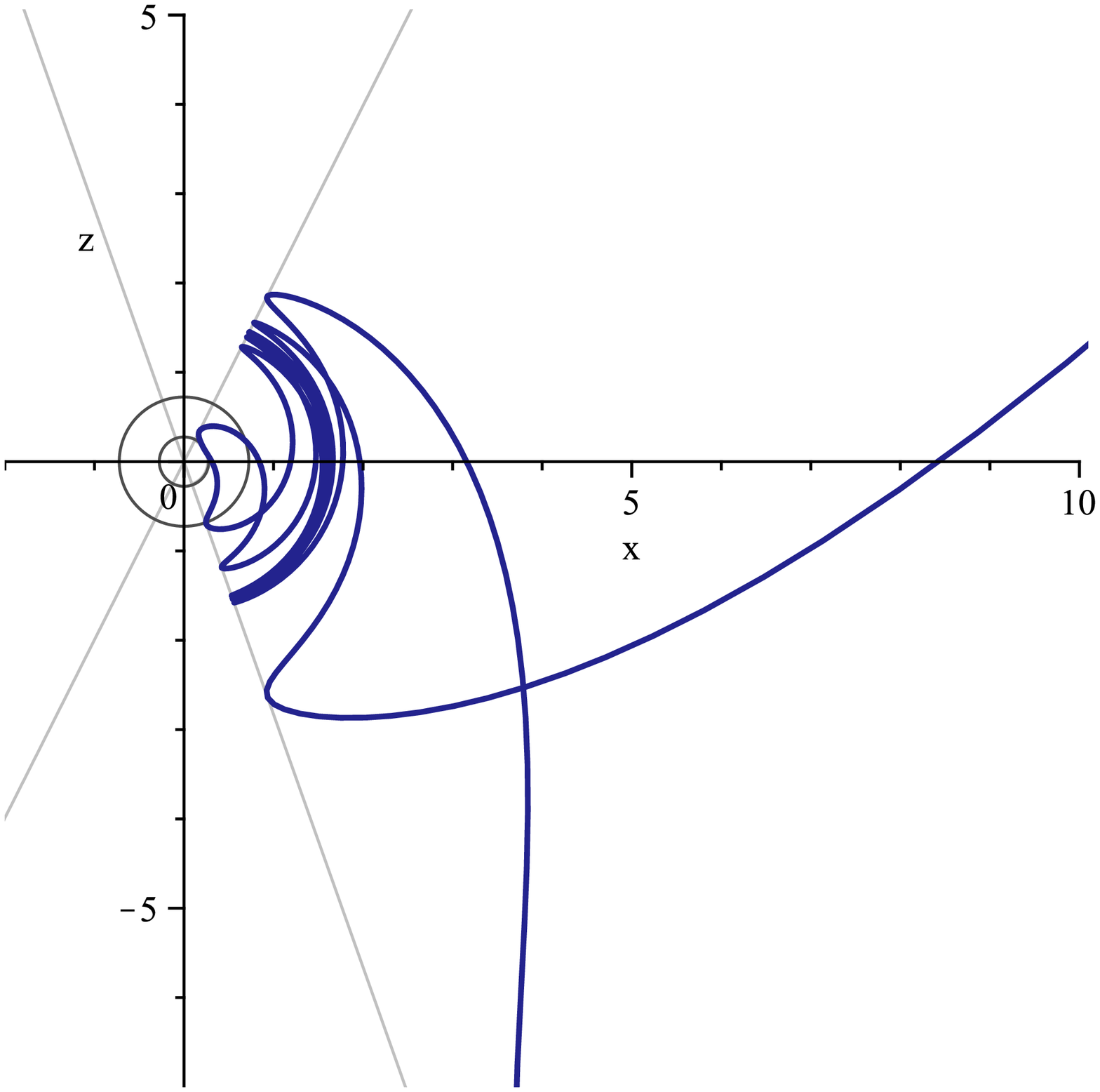}}
\caption{TEO with parameters $\tq =0.4$, $\tg =0.2$, $q=1.0$, $g=0.1$, $k=6.5$, $E=1.0039$, $\tL=1.0$. 
The particle crosses both horizons, is reflected at the potential barrier 
and emerges into a second universe,
where it moves out to infinity. 
The orbit lies on a cone with large opening angle 
$\Delta\vartheta\rightarrow\pi$.}
\label{TEO-1}
\end{figure}

\begin{figure}[htbp]
\centering
\subfigure[MBO.]{\includegraphics[width=6cm]{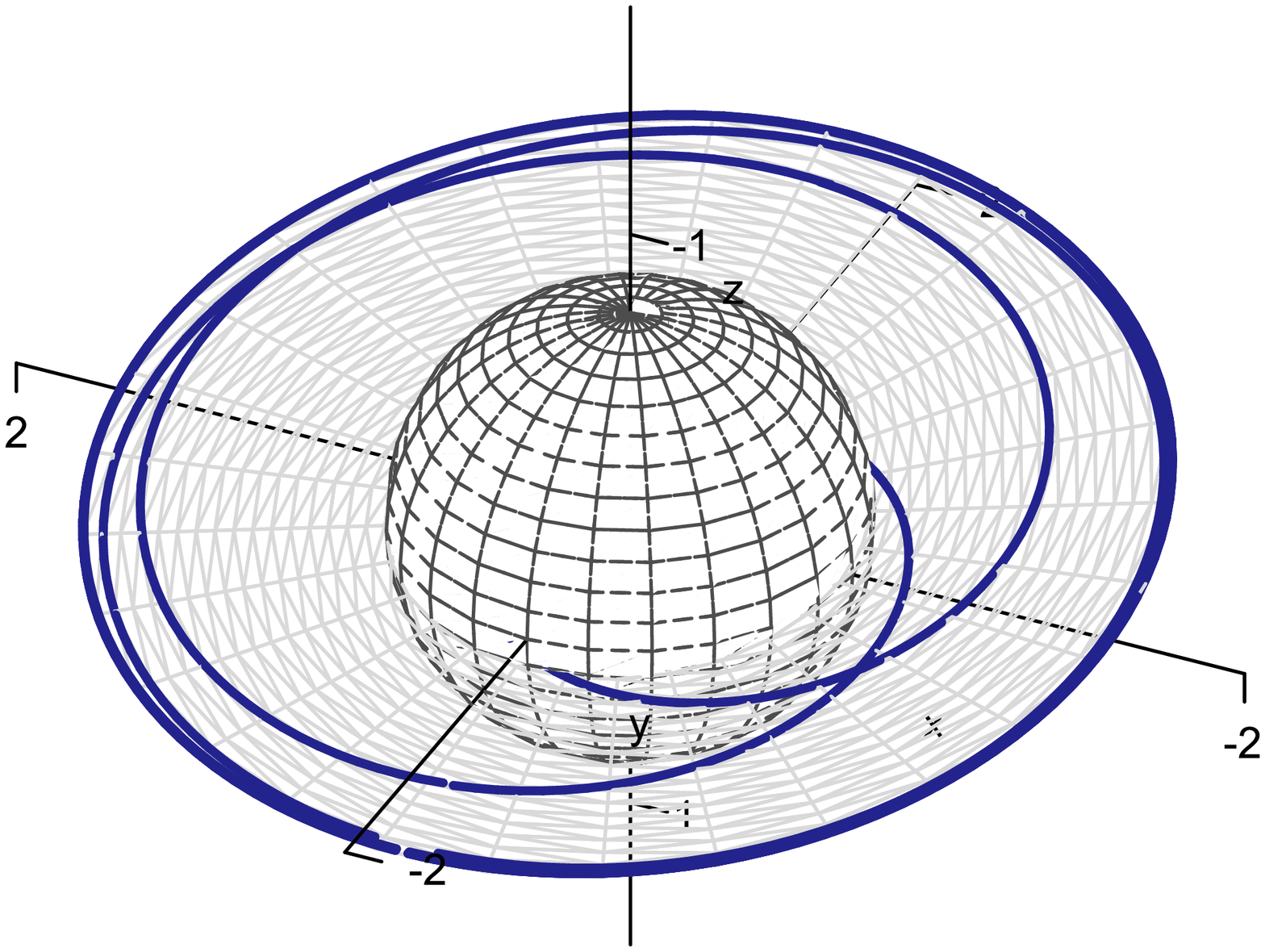}}
\subfigure[Projection onto the $x$-$z$-plane. The two circles denote 
the horizons $\tr_+$ and $\tr_-$. 
The potential barrier at $r<r_-$ prevents the particle from falling 
into the singularity.]
{\includegraphics[width=6cm]{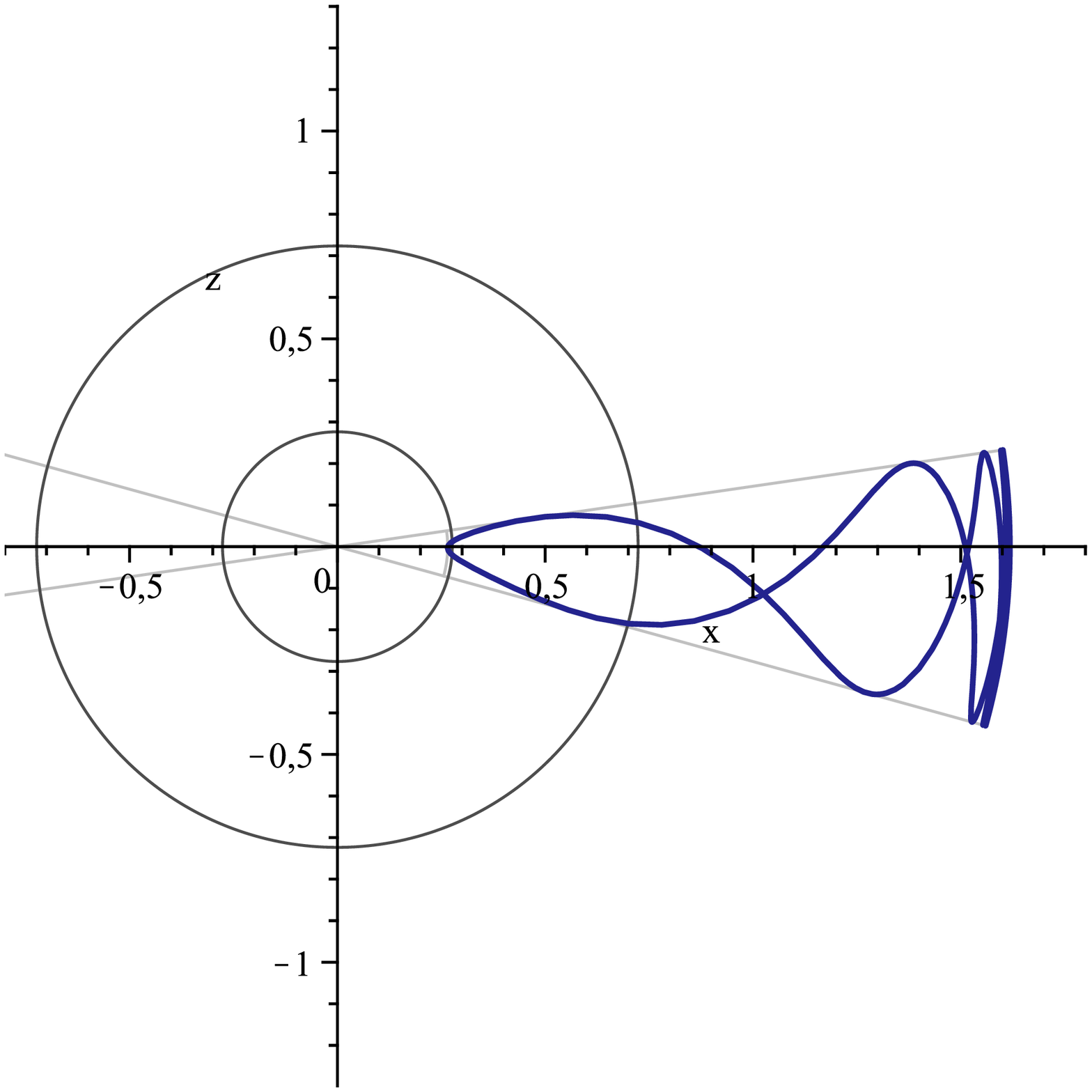}}
\subfigure[EO.]{\includegraphics[width=6cm]{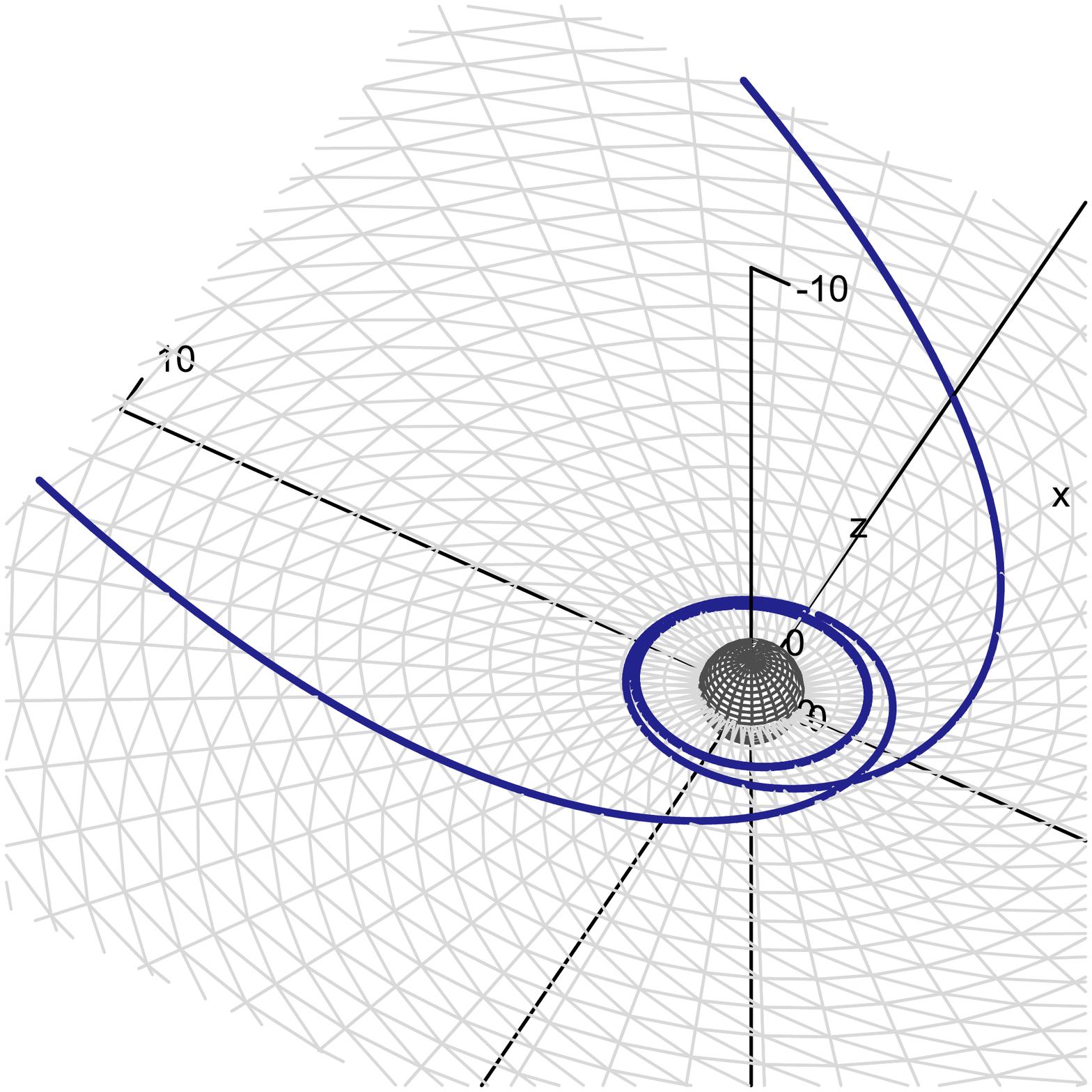}}
\subfigure[Projection onto the $x$-$z$-plane. The two circles denote $\tr_+$ and $\tr_-$.  ]{\includegraphics[width=6cm]{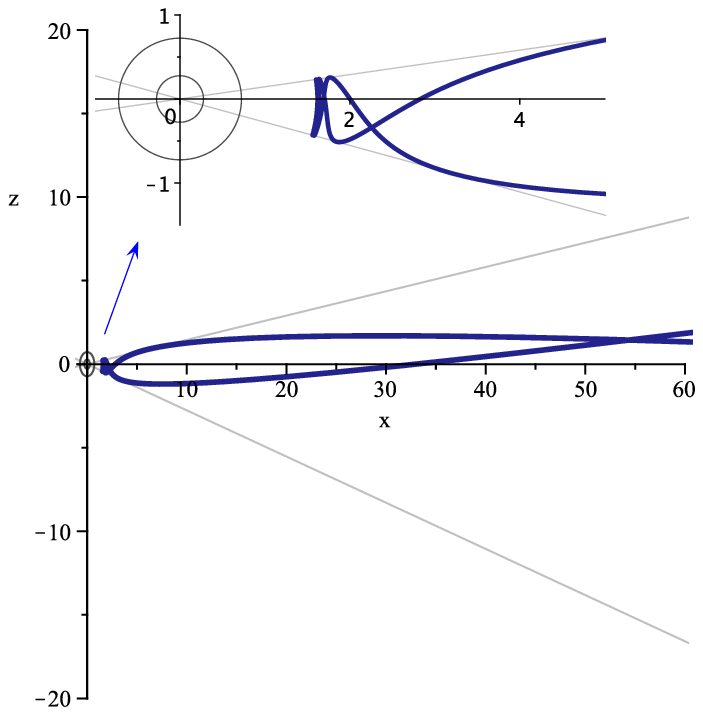}}
\caption{MBO and EO with parameters 
$\tq =0.4$, $\tg =0.2$, $q=1.0$, $g=0.1$, $k=6.5$, $E=1.0038$, $\tL=2.5$. 
An increase of $\tL$ (compare e.g.~with Fig.~\ref{TEO-1-2}) 
leads to a decrease of the opening angles of the two cones,
which confine the orbit. 
The orbits lie on cones with large opening angle 
$\Delta\vartheta\rightarrow\pi$
}
\label{MBO-2}
\end{figure}

With these analytical results we have found the complete set of orbits 
for massive charged particles. 
The following figures show solutions of the geodesic equations for the spatial coordinates. We start with small charges of a test particle in Figs.~\ref{MBO-1},~\ref{TEO-1},~\ref{MBO-2}. 
The orbits here lie on cones with large opening angle 
$\Delta\vartheta\rightarrow\pi$. 
Depending on the energy of the test particle one gets MBO, BO, TEO or EO. 
In all these figures the energy is chosen in the vicinity of the maximum 
of the effective potential. 

\begin{figure}[htbp]
\centering
\subfigure[BO. The orbit lies on a cone behind the inner horizon. 
For this orbit the singularity is not hidden. ]
{\label{BO-TEO-inside_hor1-1}\includegraphics[width=6cm]{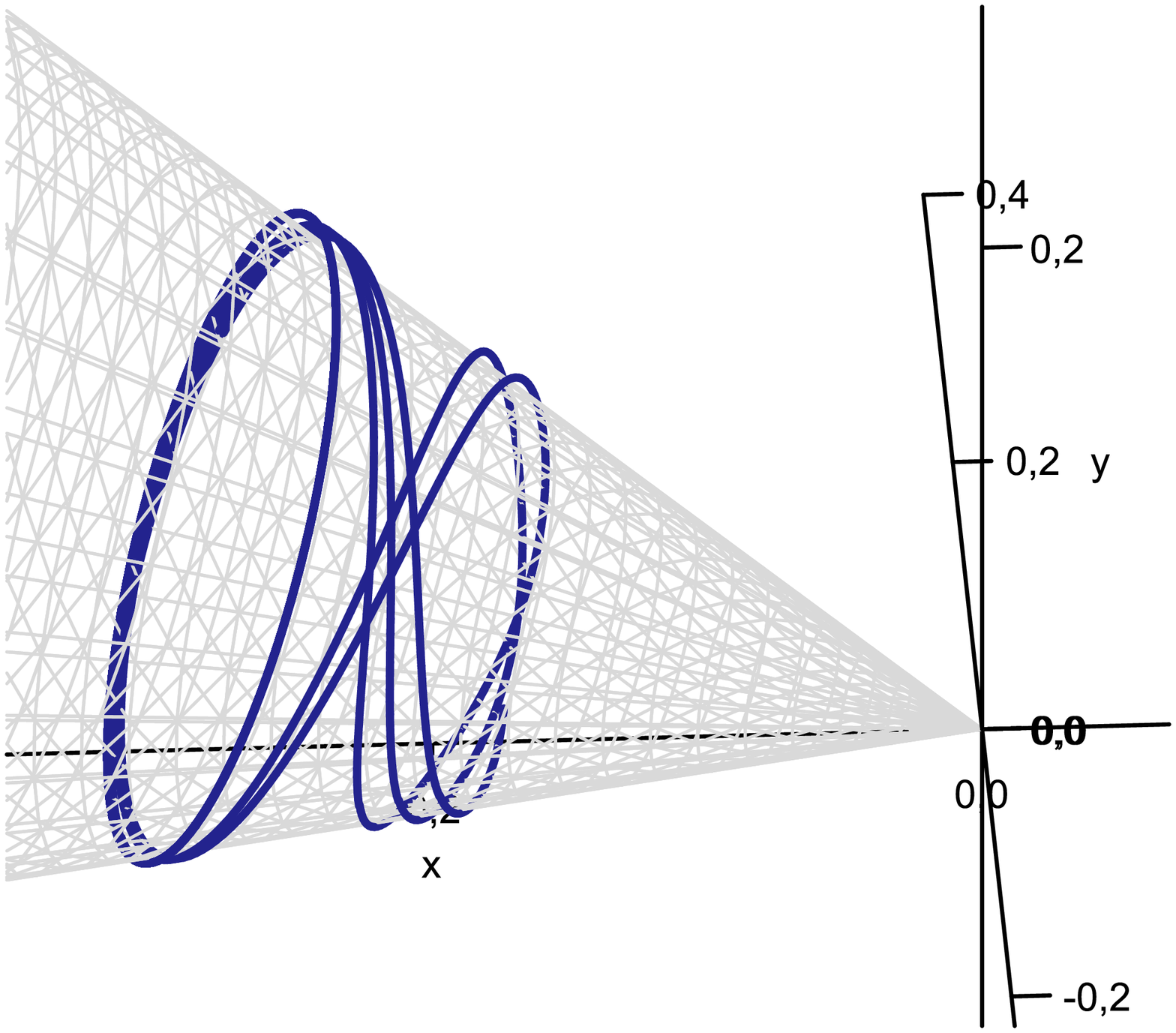}}
\subfigure[Projection onto the $x$-$z$-plane. 
The part of the black circle is the horizon $\tr_-$. 
Here $r_1, r_2 < r_-$.]{\label{BO-TEO-inside_hor1-2}\includegraphics[width=6cm]{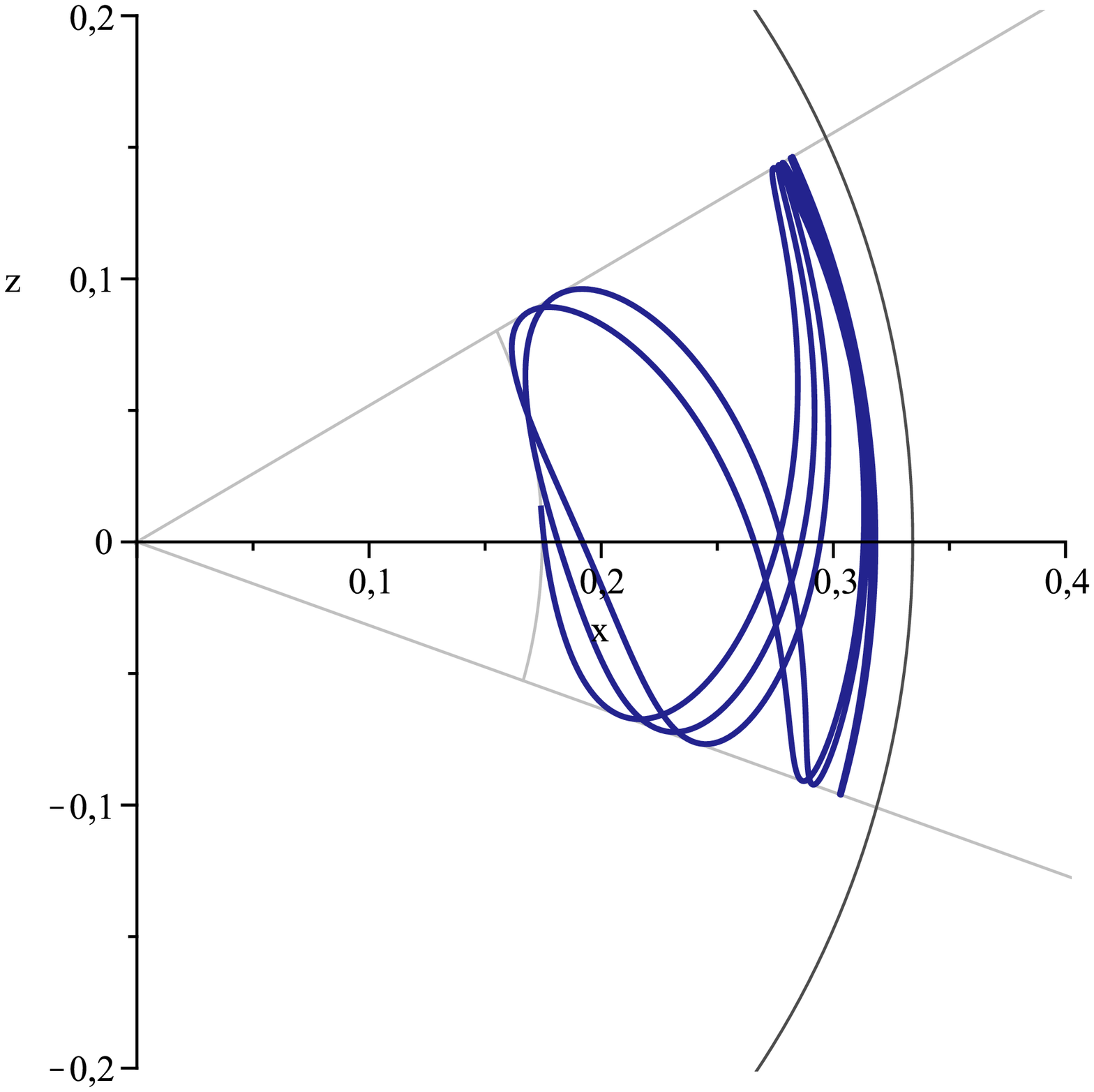}}
\subfigure[TEO]{\includegraphics[width=6cm]{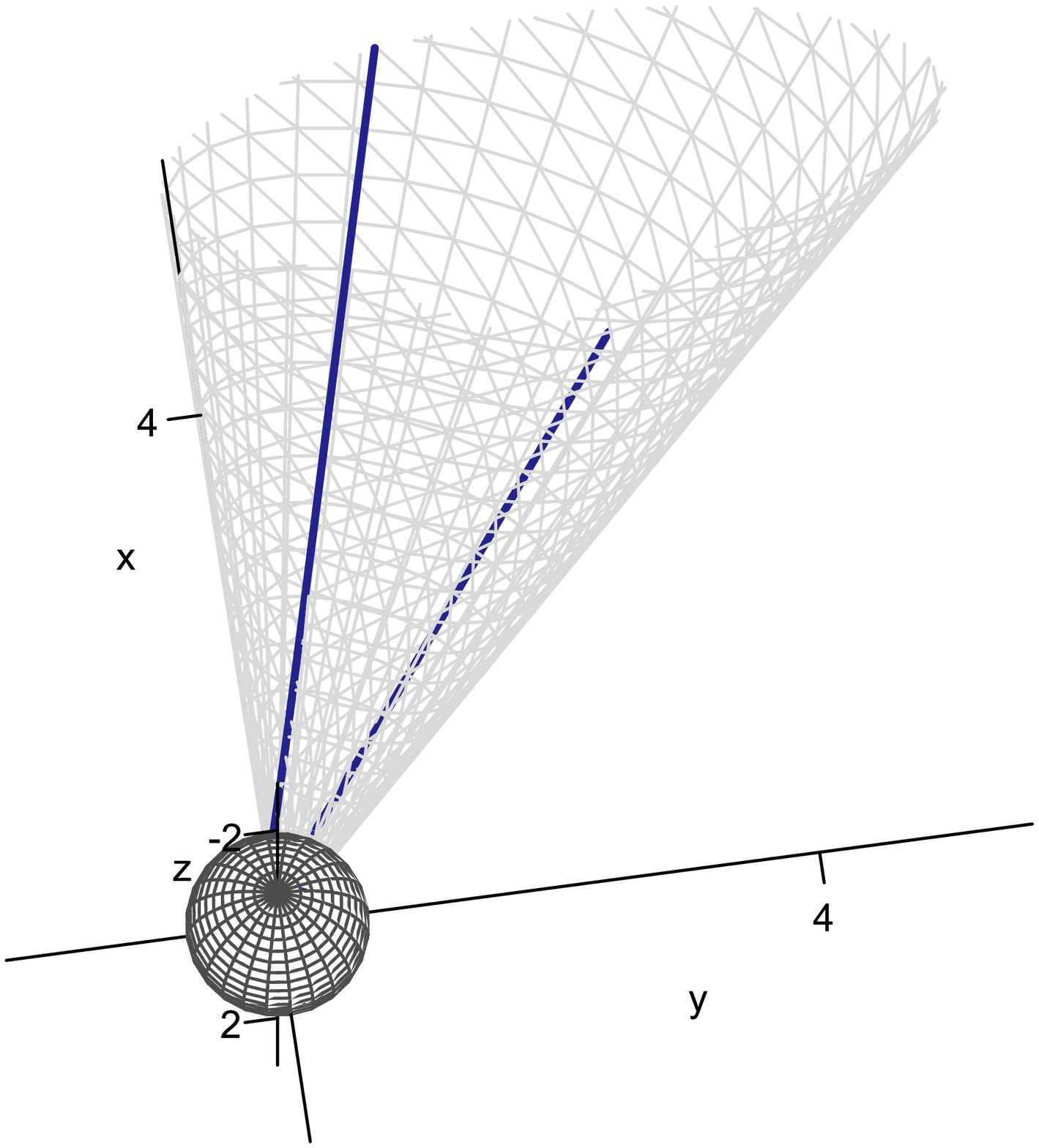}}
\subfigure[Projection onto the $x$-$z$-plane. The two circles denote 
the horizons $\tr_+$ and $\tr_-$.  ]{\includegraphics[width=6cm]{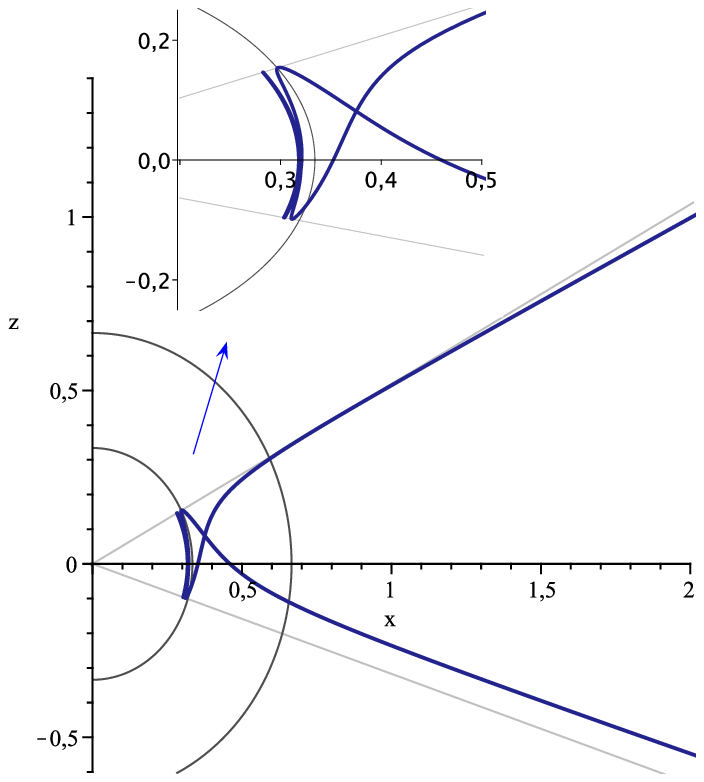}}
\caption{BO and TEO with parameters 
$\tq =0.4$, $\tg =0.25$, $q=-4$, $g=0.2$, $k=0.2$, $E=4.4673$, $\tL=0.1$. 
Because of the large charge of the test particle 
the orbital cone has a small opening angle.
}
\label{BO-TEO-inside_hor1}
\end{figure}

\begin{figure}[htbp]
\centering
\subfigure[MBO.]{\includegraphics[width=6cm]{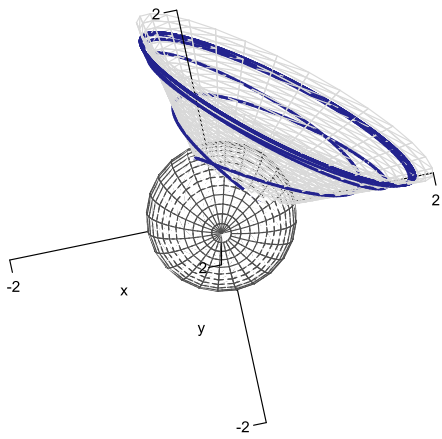}}
\subfigure[Projection onto the $x$-$z$-plane. 
The two circles denote  the horizons $\tr_+$ and $\tr_-$. 
The potential barrier at $r<r_-$ prevents the particle from falling 
into the singularity.]{\includegraphics[width=6cm]{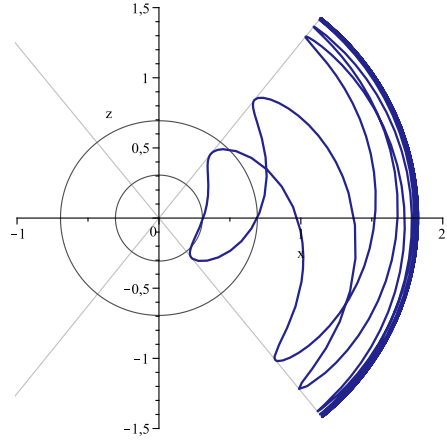}}
\subfigure[BO.]{\includegraphics[width=6cm]{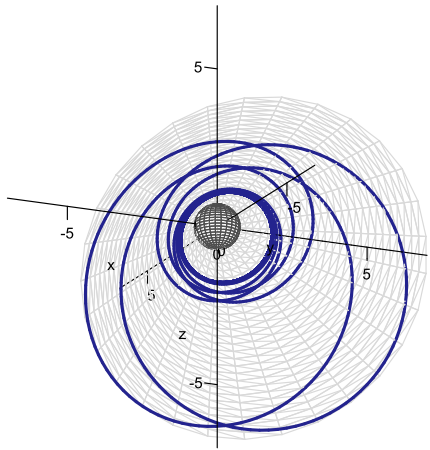}}
\subfigure[Projection onto the $x$-$z$-plane. 
The two circles denote the horizons $\tr_+$ and $\tr_-$.]{\includegraphics[width=6cm]{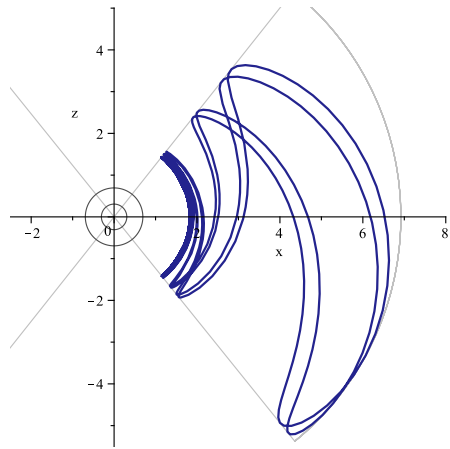}}
\caption{MBO and BO with parameters 
$\tq =0.45$, $\tg =0.1$, $q=0.0$, $g=4.0$, $k=5$, $E=0.9167$, $\tL=0.01$. 
The sphere in the 3d plots shows the horizon $\tr_+$.}
\label{MBO-oncemore}
\end{figure}

\begin{figure}[htbp]
\centering
\subfigure[TEO. The sphere shows the horizon $\tr_+$.]
{\includegraphics[width=6cm]{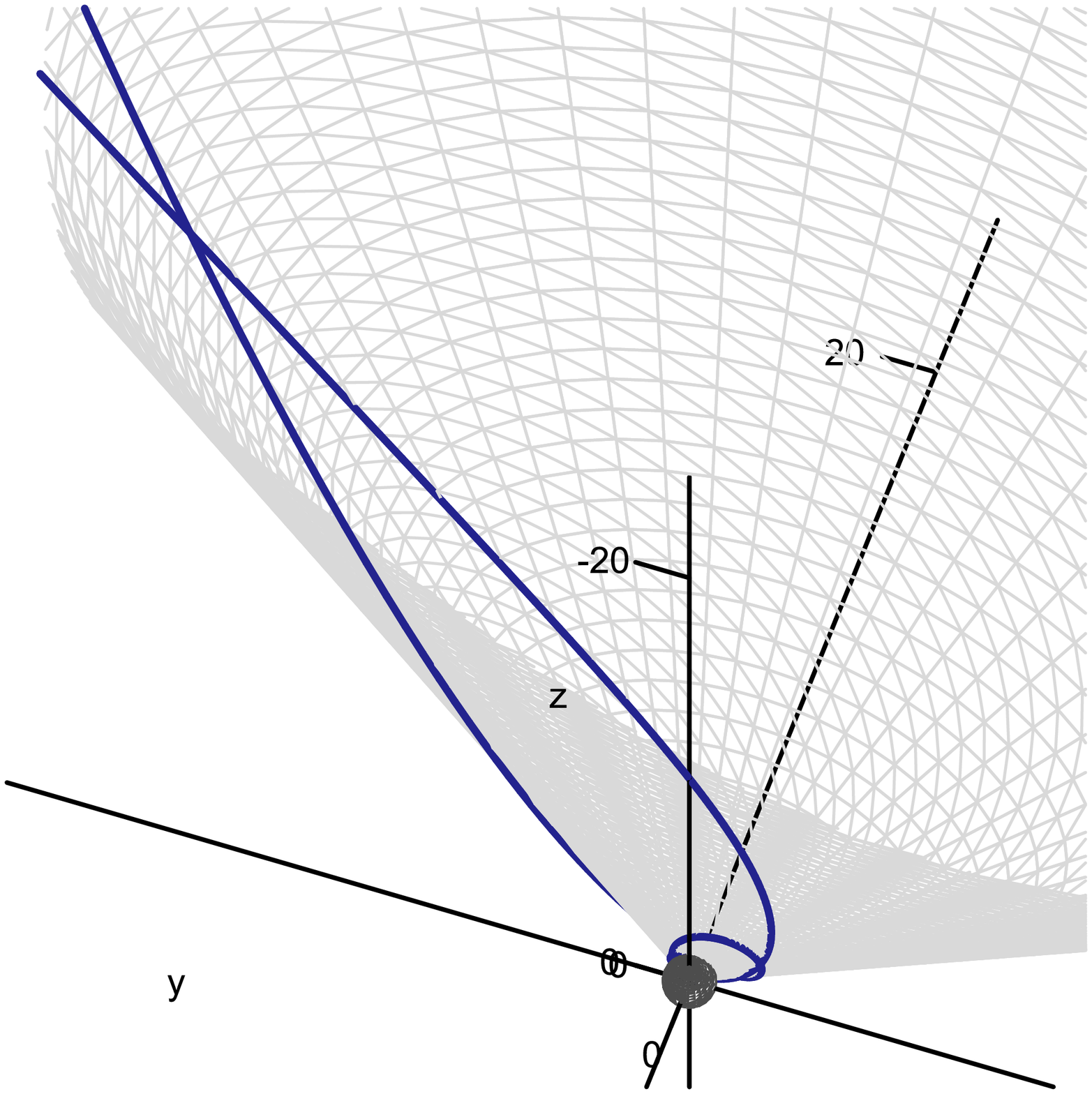}}
\subfigure[Projection onto the $x$-$z$-plane. The two circles denote 
the horizons $\tr_+$ and $\tr_-$.]{\label{TEO-1-2a}\includegraphics[width=6cm]{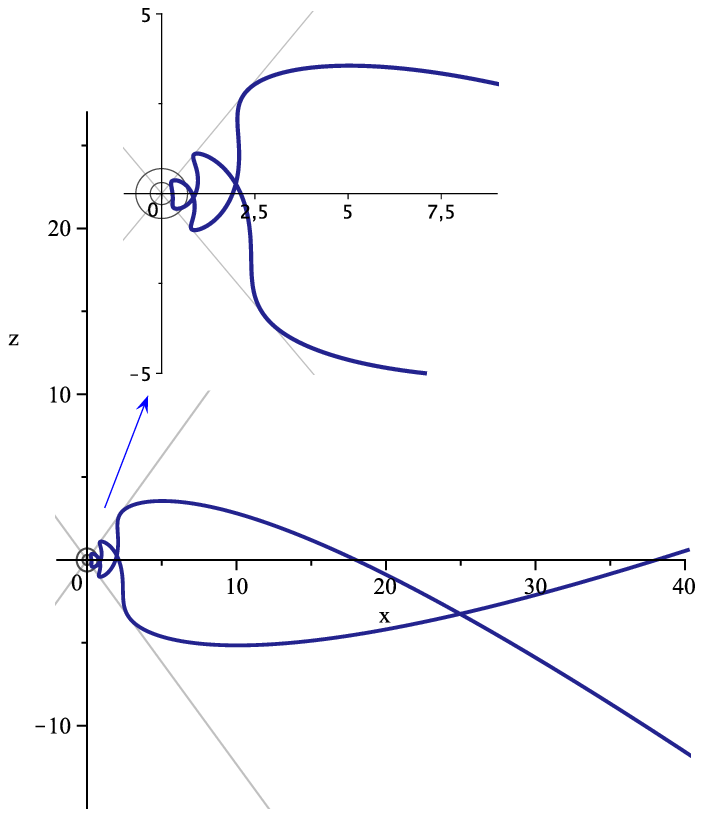}}
\caption{TEO with parameters $\tq =0.45$, $\tg =0.1$, $q=0.0$, $g=4.0$, $k=5.0$, $E=1.0001$, $\tL=0.01$. 
The particle crosses both horizons, is reflected at the potential barrier 
and emerges into a second universe, where it moves out to infinity.}
\label{TEO-2}
\end{figure}

In the Figs.~\ref{BO-TEO-inside_hor1},~\ref{MBO-oncemore},~\ref{TEO-2} 
we increased the charge of the test particle. 
The orbital cone is most evident here. 
The orbit~\ref{BO-TEO-inside_hor1-1} with its $x$-$z$ projection~\subref{BO-TEO-inside_hor1-2} is of type $\rm D$ from Table~\ref{TypesOfOrbits1}. 
The test particle orbits the singularity at $\tr=0$.
Such kinds of orbits are pecular to particles  with large charge
(and small constant ${k}$). 
An example of the effective potential is presented in Fig.~\ref{potentiale2}. 
As compared to {\it super test particles} with small mass and small charge, 
a particle with small mass and large charge is not {\it super} any more,
but it may still be regarded as a {\it test body}
or {\it test particle}~\cite{hiscock}.

\subsection{The observables}

The Reissner-Nordstr\"om space-time is believed to be of little
astrophysical relevance, because a charged black hole would 
lose its charge quickly by interactions with matter around it. 
Nevertheless, the study of charged black holes 
can help to understand more general space-times with sources 
immersed into spherically symmetric matter backgrounds. 

For test particle motion one can define frame independent observables 
such as the perihelion shift for bound orbits, 
the light deflection for escape orbit, 
the deflecton angle for flyby orbits, or the Lense--Thirring effect. 
We follow the lines of~\cite{taubnut2010} to calculate possible observables. 

We consider here bound orbits of type ${\rm E}$ and ${\rm D}$ 
from the Table~\ref{TypesOfOrbits1}. 
The $\tr$-motion is periodic with a period 
\begin{equation}
\omega_{\tr}= 2 \int_{r_{\rm min}}^{r_{\rm max}} \frac{d\tr}{\sqrt{R}} = 2 \int_{e_1}^{e_{2}} \frac{dy}{\sqrt{P_3(y)}}  \ ,
\end{equation}
where $e_1$ and $e_{2}$ are the zeros of $P_3(y)$ related to $r_{\rm min}$ and $r_{\rm max}$. The corresponding orbital frequency is $\frac{2\pi}{\omega_{\tr}}$. 

The period of the $\vartheta$--motion is given by  
\begin{equation}
\omega_{\vartheta}= 2 \int^{\vartheta_{\rm max}}_{\vartheta_{\rm min}} \frac{d\vartheta}{\sqrt{\Theta}}= - 2 \int^{\xi_{\rm min}}_{\xi_{\rm max}} \frac{d\xi}{\sqrt{\Theta_{\xi}}} = \frac{2\pi}{\sqrt{-a}} \ ,
\end{equation}
and the corresponding frequency by $\frac{2\pi}{\omega_{\vartheta}}$.

The secular accumulation rates of the angle $\varphi$ and the time $t$ are given by: 
\begin{equation}
Y_{\varphi} = \frac{2}{\omega_{\vartheta}} \int^{\xi_{{\rm min}}}_{\xi_{{\rm max}}} \frac{\tilde{L} + \Dg \xi}{1 - \xi^2} \left(-\frac{d\xi}{\sqrt{\Theta_\xi}} \right)= \frac{1}{\omega_\vartheta} (I_+ + I_-)\bigr|^{\xi_{\rm min}}_{\xi_{\rm max}} = \sqrt{-a} \, . 
\end{equation}
and
\begin{eqnarray}
\Gamma & = & \frac{2}{\omega_{\tr}} \left(  E \int_{r_{\rm min}}^{r_{\rm max}}{ \frac{\tr^4}{\tDr} \frac{d\tr}{\sqrt{R}}} 
+ \De \int_{r_{\rm min}}^{r_{\rm max}}{  \frac{\tr^3}{\tDr} \frac{d\tr}{\sqrt{R}}} \right) \nonumber \\ 
 &=& \frac{2}{\omega_{\tr}} \left(  I^1_{\tr}\Bigl|^{\gamma_{e_2} }_{\gamma_{e_1} } + I^2_{\tr}\Bigl|^{\gamma_{e_2} }_{\gamma_{e_1} } \right) \ ,
\end{eqnarray}
where $I^1_{\tr}$ and $I^2_{\tr}$ defined in the equations~\eqref{IryNUT3} and~\eqref{IryNUT_I2} are evaluated at $\gamma_{e_i}$ corresponding to the root $e_i$, $i=1,2$. The orbital frequences $\Omega_r$, $\Omega_\vartheta$ and $\Omega_\varphi$ are then given by:
\begin{equation}
\Omega_{\tr}=\frac{2\pi}{\omega_{\tr}}\frac{1}{\Gamma} \, , \qquad \Omega_\vartheta=\frac{2\pi}{\omega_{\vartheta}}\frac{1}{\Gamma} \, , \qquad \Omega_\varphi=\frac{Y_{\varphi}}{\Gamma} \ .
\end{equation}
The perihelion shift and the Lense--Thirring effects are defined as differences between these orbital frequences
\begin{eqnarray}
\Delta_{\rm perihelion} & = & \Omega_\varphi - \Omega_{\tr} = \left( \sqrt{-a} - \frac{2\pi}{\omega_{\tr}}  \right)\frac{1}{\Gamma} \\ 
\Delta_{\rm Lense-Thirring} & = & \Omega_\varphi - \Omega_\vartheta = 0 \ .
\end{eqnarray}
As expected, there is no Lense--Thirring effect.

\section{Conclusions and Outlook}

In this paper we presented the analytic solution of the geodesic equations of charged test particles in the Reissner-Nordstr\"om space-time 
in terms of the Weierstra{\ss} $\wp$, $\sigma$ and $\zeta$ functions. The derived orbits depend on the particle's energy, angular momentum, electric and magnetic charges, Carter constant and on the parameters of the gravitating and charged source. We discussed the general structure of the orbits and gave a complete classification of their types. The orbits of charged test particles lie on a cone similar to the motion in Taub-NUT space-times with gravitomagnetic charge. This happens because of the non-vaning $\Dg\cos\vartheta$ term in the equations~\eqref{eq-r-theta:2} and~\eqref{dvarphidgamma}. 
Thus, the motion of only electrically charged particles around 
an only electrically charged source 
or the motion of only magnetically charged particles around
an only magnetically charged source 
reduces to motion in a plane. 
However, this is not true for the motion of test particles with both types of
charge around a charged source
or the motion of charged particles around a source with both types of
charges (as in the examples considered in this paper). 

We showed that for large charge of test particles bound orbits 
behind the inner horizon are possible. 
So, a test particle moves in a space-time,
that has horizons which hide the singularity, 
but for the motion of the particle the singularity is not hidden. 
Such an effect is not present for neutral particles. 

Finally we remark, that as compared to the geodesics
in the field of a gravitomagnetic monopole~\cite{taubnut2010}, 
the geodesics in the Reissner-Norstr\"om space-time are complete,
because the space-time can be the analytically continued
(without destroying some of its essential properties~\cite{MiKruGo71}). 
Thus at least theoretically,
a particle crossing the horizons may emerge afterwards 
into another universe, following a many-world orbit.

\section*{Acknowledgement}

We would like to thank Jutta Kunz and Claus L\"ammerzahl 
for suggesting this research topic and for
helpful discussions and interesting remarks. 
We also acknowledge fruitful discussions with Eva Hackmann. 
V.K.~acknowledges financial support of the German Research Foundation DFG. 

\begin{appendix}

\section{Integration of elliptic integrals of the third kind}\label{elldiffIII}

We consider an integral of the type 
$\displaystyle{I_1=\int^v_{v_{\rm in}} \frac{dv}{\wp(v)-p_1}}$. 
This integral is of the third kind because the function $f_1(v)=\left(\wp(v)-p_1\right)^{-1}$ has two simple poles $v_1$ and $v_2$ in a fundamental parallelogram with vertices $0$, $2\omega_1$, $2\omega_1+2\omega_2$, $2\omega_2$, where $2\omega_1$ and $2\omega_2$ are fundamental periods of $\wp(v)$ and $\wp^\prime(v)$. 

Consider the Laurent series for the function $f_1$ around $v_i$
\begin{equation}
f_1(v)= a_{-1,i} (v-v_i)^{-1} + \text{holomorphic part}  \ , \label{f1Laurent}
\end{equation} 
and the Taylor series of $f_1^{-1}$ about $v_i$
\begin{equation}
f^{-1}_1(v)= \wp^\prime(v_i)(v-v_i) + {\cal{O}}(v^2)  \ . \label{f1Taylor}
\end{equation} 
Comparing the coefficients in the equality $1=f_1(v)f^{-1}_1(v)$ where
\begin{equation}
1 = f_1(v)\left(  \wp^\prime(v_i)(v-v_i) + {\cal{O}}(v^2)  \right) =  a_{-1,i} \wp^\prime(v_i) + {\cal{O}}(v^2)   \, \label{f1coeff}
\end{equation} 
yields $\displaystyle{a_{-1,i}=\frac{1}{\wp^\prime(v_i)}}$. Thus, the function $f_1(v)$ has a residue $\frac{1}{\wp^\prime(v_i)}$ in $v_i$.

The Weierstra{\ss} $\zeta(v)$ function is an elliptic function with a simple pole in $0$ and residue $1$. Then the function $A_1=f_1(v)-\sum^{2}_{i=1}\frac{\zeta(v-v_i)}{\wp^\prime(v_i)}$ is an elliptic function without poles and therefore a constant~\cite{Markush}, which can be determined from $f_1(0)=0$. Thus,
\begin{equation}
f_1(v)= \sum^{2}_{i=1}\frac{\zeta(v-v_i)+\zeta(v_i)}{\wp^\prime(v_i)}  \ , \label{f1fin}
\end{equation} 
here $\wp^\prime(v_2)=\wp^\prime(2\omega_j-v_1)=-\wp^\prime(v_1)$.  
Applying now the definition of the Weierstra{\ss} $\sigma$--function $\int^{v}_{v_{\rm in}} \zeta(v)dv=\log\sigma(v)-\log\sigma(v_{\rm in})$ upon the integral $I_1$ we get the solution
\begin{equation}
I_1= \int^v_{v_{\rm in}} f_1(v)dv=\sum^{2}_{i=1}\frac{1}{\wp^\prime(v_i)} \Biggl( \zeta(v_i)(v-v_{\rm in}) + \log\frac{\sigma(v-v_i)}{\sigma(v_{\rm in}-v_i)}  \Biggr)  \ . \label{I1int}
\end{equation}

\section{Integration of elliptic integrals of the type $I_2=\int^v_{v_{\rm in}} \frac{dv}{\left(\wp(v)-p_3 \right)^2}$ }\label{elldiff2}

We consider the Laurent series of $f_2(v)$ and the Taylor series of $f^{-1}_2(v)$ around $v_i$ for $\displaystyle{f_2(v)=\frac{1}{\left(\wp(v)-p_3 \right)^2}}$:
\begin{eqnarray}
  f_2(v) &=& a_{-2,i} (v-v_i)^{-2} + a_{-1,i} (v-v_i)^{-1} + \text{holomorphic part}  \ , \label{f2Laurent} \\
  f^{-1}_2(v) &=& \left( \wp^\prime(v_i)(v-v_i) + \half \wp^{\prime\prime}(v_i) (v-v_i)^2 + {\cal{O}}(v^3) \right)^2 \nonumber \\  & =& \left(\wp^\prime(v_i)\right)^2(v-v_i)^2 + \wp^\prime(v_i) \wp^{\prime\prime}(v_i) (v-v_i)^3 + {\cal{O}}(v^4) \label{f2Taylor} \ . 
\end{eqnarray}
The function $f_2(v)$ has poles of second order in $v_1$ and $v_2$ such that $f_2(v_1)=p_3=f_2(v_2)$. 
Comparison of the coefficients in 
\begin{eqnarray}
1 & = & f_2(v)\left( \left(\wp^\prime(v_i)\right)^2(v-v_i)^2 + \wp^\prime(v_i) \wp^{\prime\prime}(v_i) (v-v_i)^3 + {\cal{O}}(v^4)   \right) \nonumber \\
  & = & a_{-2,i} \left(\wp^\prime(v_i)\right)^2 + (v-v_i) \left[ a_{-1,i}\left(\wp^\prime(v_i)\right)^2 +  a_{-2,i}\wp^\prime(v_i)\wp^{\prime\prime}(v_i)    \right] +  {\cal{O}}(v^2) \label{f2coeff}  \, 
\end{eqnarray} 
yields
\begin{equation}
a_{-2,i}=\frac{1}{\left(\wp^\prime(v_i)\right)^2} \,\, \ , \,\,\,\,\,\,\, a_{-1,i}=-\frac{\wp^{\prime\prime}(v_i)}{\left(\wp^\prime(v_i)\right)^3}  \ . \label{a2a1f2}
\end{equation}

The function $\wp(v)$ possesses a pole of second order in $v=0$ with residuum $0$ and the Laurent series of $\wp$ begins with $v^{-2}$. Then the Laurent series of $a_{-2,i}\wp(v-v_i)$ around $v_i$ begins with $a_{-2,i}(v-v_i)^{-2}$ which is similar to the first term in~\eqref{f2Laurent}. The Laurent series of $a_{-1,i}\zeta(v-v_i)$ around $v_i$ begins with $a_{-1,i}(v-v_i)^{-1}$. Thus, the function $\displaystyle{A_2=f_2(v)-\sum^{2}_{i=1}\left(  \frac{\wp(v-v_i)}{\left(\wp^\prime(v_i)\right)^2} - \frac{\wp^{\prime\prime}(v_i)\zeta(v-v_i)}{\left(\wp^\prime(v_i)\right)^3}  \right)}$ has no poles and is constant~\cite{Markush} and can be calculated from $f_2(0)=0$: $\displaystyle{A_2=-\sum^{2}_{i=1}\left(  \frac{\wp(v_i)}{\left(\wp^\prime(v_i)\right)^2} + \frac{\wp^{\prime\prime}(v_i)\zeta(v_i)}{\left(\wp^\prime(v_i)\right)^3}  \right)}$.

Using of $\int^v_{v_{\rm in}}\wp(v)dv = - \zeta(v) + \zeta(v_{\rm in})$ and the definition of the $\sigma$--function the integral $I_2$ takes the form:
\begin{equation}
I_2 = \int^v_{v_{\rm in}} f_2(v)dv= A_2 (v-v_{\rm in})  
- \sum^{2}_{i=1} \Biggl[ \zeta(v-v_i) - \zeta(v_{\rm in}-v_i) + \frac{\wp^{\prime\prime}(v_i)}{\wp^\prime(v_i)}
 \log\frac{\sigma(v-v_i)}{\sigma(v_{\rm in}-v_i)} \Biggr]\frac{1}{\left(\wp^\prime(v_i)\right)^2}   \ . \label{I2int}
\end{equation}

\end{appendix}


\bibliographystyle{unsrt}

\end{document}